\documentclass[12pt,preprint]{aastex}
\usepackage{times,pslatex}
\usepackage{amsmath,amsfonts,graphicx}
\usepackage{epsfig}
\usepackage{rotate}
\usepackage{longtable}

\def\kms{\ifmmode{\rm km\thinspace s^{-1}}\else km\thinspace s$^{-1}$\fi}

\newcommand{\vRstarA}{1.7903}
\newcommand{\eRstarA}{0.0055}
\newcommand{\vRstarB}{0.9663}
\newcommand{\eRstarB}{0.0057}
\newcommand{\vRplanetear}{11.8739}
\newcommand{\eRplanetear}{0.1377}
\newcommand{\vRplanetjup}{1.06}
\newcommand{\eRplanetjup}{0.01}
\newcommand{\vaplanet}{2.7205}
\newcommand{\eaplanet}{0.0070}
\newcommand{\vebin}{0.1602}
\newcommand{\eebin}{0.0004}
\newcommand{\vabin}{0.1276}
\newcommand{\eabin}{0.0002}

\newcommand{\vMstarA}{1.2207}
\newcommand{\eMstarA}{0.0112}
\newcommand{\vMstarB}{0.9678}
\newcommand{\eMstarB}{0.0039}
\newcommand{\vMplanetear}{483}
\newcommand{\eMplanetear}{206}
\newcommand{\vMplanetjup}{1.52}
\newcommand{\eMplanetjup}{0.65}

\newcommand{\kepler}{{\it Kepler}~}
\newcommand{\keplerp}{{\it Kepler}}
\newcommand{\kic}{{Kepler-1647~}}
\newcommand{\kiccomma}{{Kepler-1647}}
\newcommand{\kicb}{{Kepler-1647b~}}
\newcommand{\kicbcomma}{{Kepler-1647b}}

\newcommand{\Mjup}{\mbox{$M_{\rm Jup}$}}
\newcommand{\Rjup}{\mbox{$R_{\rm Jup}$}}

\newcommand{\Msun}{\mbox{$M_\odot$}}
\newcommand{\Rsun}{\mbox{$R_\odot$}}

\slugcomment{Accepted for publication in the Astrophysical Journal}
\shortauthors{Kostov et al.}

\received{2015, Nov}
\begin{document}

\title{\kicbcomma: the largest and longest-period \kepler transiting circumbinary planet.}
\shorttitle{Circumbinary Planet \kicb}

\author{
Veselin~B.~Kostov\altaffilmark{1, 25},
Jerome~A.~Orosz\altaffilmark{2},
William~F.~Welsh\altaffilmark{2},
Laurance~R.~Doyle\altaffilmark{3},
Daniel~C.~Fabrycky\altaffilmark{4},
Nader~Haghighipour\altaffilmark{5},
Billy~Quarles\altaffilmark{6, 7, 25},
Donald~R.~Short\altaffilmark{2},
William~D.~Cochran\altaffilmark{8}, 
Michael~Endl\altaffilmark{8},
Eric~B.~Ford\altaffilmark{9},
Joao~Gregorio\altaffilmark{10},
Tobias~C. Hinse\altaffilmark{11,12},
Howard~Isaacson\altaffilmark{13},
Jon~M.~Jenkins\altaffilmark{14},
Eric~L.~N.~Jensen\altaffilmark{15},
Stephen Kane\altaffilmark{16}, 
Ilya~Kull\altaffilmark{17},
David~W.~Latham\altaffilmark{18},
Jack~J.~Lissauer\altaffilmark{14},
Geoffrey~W.~Marcy\altaffilmark{13},
Tsevi~Mazeh\altaffilmark{17},
Tobias~W.~A.~M\"{u}ller\altaffilmark{19},
Joshua~Pepper\altaffilmark{20},
Samuel~N.~Quinn\altaffilmark{21, 26},
Darin~Ragozzine\altaffilmark{22},
Avi~Shporer\altaffilmark{23,27},
Jason~H.~Steffen\altaffilmark{24},
Guillermo~Torres\altaffilmark{18},
Gur~Windmiller\altaffilmark{2},
William~J.~Borucki\altaffilmark{14}
}

\email{veselin.b.kostov@nasa.gov}

\altaffiltext{1}{NASA Goddard Space Flight Center, Mail Code 665, Greenbelt, MD, 20771}
\altaffiltext{2}{Department of Astronomy, San Diego State University, 5500 Campanile Drive, San Diego, CA 92182}
\altaffiltext{3}{SETI Institute, 189 Bernardo Avenue, Mountain View, CA 94043; and Principia College, IMoP, One Maybeck Place, Elsah, Illinois 62028}
\altaffiltext{4}{Department of Astronomy and Astrophysics, University of Chicago, 5640 South Ellis Avenue, Chicago, IL 60637}
\altaffiltext{5}{Institute for Astronomy, University of Hawaii-Manoa, Honolulu, HI 96822, USA}
\altaffiltext{6}{Department of Physics and Physical Science, The University of Nebraska at Kearney, Kearney, NE 68849}
\altaffiltext{7}{NASA Ames Research Center, Space Science Division MS 245-3, Code SST, Moffett Field, CA 94035}
\altaffiltext{8}{McDonald Observatory, The University of Texas as Austin, Austin, TX 78712-0259}
\altaffiltext{9}{Department of Astronomy and Astrophysics, The Pennsylvania State University, 428A Davey Lab, University Park, PA 16802, USA}
\altaffiltext{10}{Atalaia Group and Crow-Observatory, Portalegre, Portugal}
\altaffiltext{11}{Korea Astronomy and Space Science Institute (KASI), Advanced Astronomy and Space Science Division, Daejeon 305-348, Republic of Korea}
\altaffiltext{12}{Armagh Observatory, College Hill, BT61 9DG Armagh, Northern Ireland, UK}
\altaffiltext{13}{Department of Astronomy, University of California Berkeley, 501 Campbell Hall, Berkeley, CA 94720, USA}
\altaffiltext{14}{NASA Ames Research Center, Moffett Field, CA 94035, USA}
\altaffiltext{15}{Department of Physics and Astronomy, Swarthmore College, Swarthmore, PA 19081, USA}
\altaffiltext{16}{Department of Physics and Astronomy, San Francisco State University, 1600 Holloway Avenue, San Francisco, CA 94132, USA}
\altaffiltext{17}{Department of Astronomy and Astrophysics, Tel Aviv University, 69978 Tel Aviv, Israel}
\altaffiltext{18}{Harvard-Smithsonian Center for Astrophysics, 60 Garden Street, Cambridge, MA 02138}
\altaffiltext{19}{Institute for Astronomy and Astrophysics, University of Tuebingen, Auf der Morgenstelle 10, D-72076 Tuebingen, Germany}
\altaffiltext{20}{Department of Physics, Lehigh University, Bethlehem, PA 18015, USA}
\altaffiltext{21}{Department of Physics and Astronomy, Georgia State University, 25 Park Place NE Suite 600, Atlanta, GA 30303}
\altaffiltext{22}{Department of Physics and Space Sciences, Florida Institute of Technology, 150 W. University Blvd, Melbourne, FL 32901, USA}
\altaffiltext{23}{Jet Propulsion Laboratory, California Institute of Technology, 4800 Oak Grove Drive, Pasadena, CA 91109, USA}
\altaffiltext{24}{Department of Physics and Astronomy, Northwestern University, 2145 Sheridan Road, Evanston, IL 60208}
\altaffiltext{25}{NASA Postdoctoral Fellow}
\altaffiltext{26}{NSF Graduate Research Fellow}
\altaffiltext{27}{Sagan Fellow}

\begin{abstract}
We report the discovery of a new \kepler transiting circumbinary planet (CBP). This latest addition to the still-small family of CBPs defies the current trend of known short-period planets orbiting near the stability limit of binary stars. Unlike the previous discoveries, the planet revolving around the eclipsing binary system \kic has a very long orbital period ($\sim1100$ days) and was at conjunction only twice during the \kepler mission lifetime. Due to the singular configuration of the system, \kicb is not only the longest-period transiting CBP at the time of writing, but also one of the longest-period transiting planets. With a radius of $\vRplanetjup\pm\eRplanetjup~\Rjup$ it is also the largest CBP to date. The planet produced three transits in the light-curve of \kic (one of them during an eclipse, creating a syzygy) and measurably perturbed the times of the stellar eclipses, allowing us to measure its mass to be $\vMplanetjup\pm\eMplanetjup~\Mjup$. The planet revolves around an 11-day period eclipsing binary consisting of two Solar-mass stars on a slightly inclined, mildly eccentric ($e_{bin} = 0.16$), spin-synchronized orbit. Despite having an orbital period three times longer than Earth's, \kicb is in the conservative habitable zone of the binary star throughout its orbit.
\end{abstract}

\keywords{binaries: eclipsing -- planetary systems -- stars: individual
(\object{KIC-5473556, KOI-2939, Kepler-1647}) -- techniques: photometric}

\section{Introduction}
\label{sec:intro}

Planets with more than one sun have long captivated our collective imagination, yet direct evidence of their existence has emerged only in the past few years. Eclipsing binaries, in particular, have long been thought of as ideal targets to search for such planets (Borucki \& Summers~1984, Schneider \& Chevreton~1990, Schneider \& Doyle~1995, Jenkins et al.~1996, Deeg et al.~1998). Early efforts to detect transits of a circumbinary planet around the eclipsing binary system CM Draconis---a particularly well-suited system composed of two M-dwarfs on a nearly edge-on orbit---suffered from incomplete temporal coverage (Schneider \& Doyle~1995, Doyle et al.~2000, Doyle \& Deeg~2004). Several non-transiting circumbinary candidates have been proposed since 2003, based on measured timing variations in binary stellar systems (e.g.,~Zorotovic \& Schreiber 2013). The true nature of these candidates remains, however, uncertain and rigorous dynamical analysis has challenged the stability of some of the proposed systems (e.g.,~Hinse et al.~2014, Schleicher et al.~2015). It was not until 2011 and the continuous monitoring of thousands of eclipsing binaries (hereafter EBs) provided by NASA's {\it Kepler} mission that the first circumbinary planet, \keplerp-16b, was unambiguously detected through its transits (Doyle et al.~2011). Today, data from the mission has allowed us to confirm the existence of 10 transiting circumbinary planets in 8 eclipsing binary systems (Doyle et al.~2011, Welsh et al.~2012, 2015, Orosz et al.~2012ab, 2015, Kostov et al.~2013, 2014, Schwamb et al.~2013). Curiously enough, these planets have all been found to orbit EBs from the long-period part of the \kepler EB distribution, and have orbits near the critical orbital separation for dynamical stability (Welsh et al.~2014, Martin et al.~2015). 

These exciting new discoveries provide better understanding of the formation and evolution of planets in multiple stellar system, and deliver key observational tests for theoretical predictions (e.g.,~Paardekooper et al.~2012, Rafikov~2013, Marzari et al.~2013, Pelupessy \& Portegies Zwart~2013, Meschiari~2014, Bromley \& Kenyon~2015, Silsbee \& Rafikov~2015, Lines et al.~2015, Chavez et al.~2015, Kley \& Haghighipour~2014, 2015, Miranda \& Lai~2015). Specifically, numerical simulations indicate that CBPs should be common, typically smaller than Jupiter, and close to the critical limit for dynamical stability -- due to orbital migration of the planet towards the edge of the precursor disk cavity surrounding the binary star (Pierens \& Nelson, 2007, 2008, 2015, hereafter PN07, PN08, PN15). Additionally, the planets should be co-planar (within a few degrees) for binary stars with sub-AU separation due to disk-binary alignment on precession timescales (Foucart \& Lai~2013,~2014). The orbital separation of each new CBP discovery, for example, constrains the models of protoplanetary disks and migration history and allow us to discern between an observational bias or a migration pile-up (Kley \& Haghighipour~2014, 2015). Discoveries of misaligned transiting CBPs such as \keplerp-413b (Kostov et al.~2014) and \keplerp-453b (Welsh et al.~2015) help determine the occurrence frequency of CBPs by arguing for the inclusion of a distribution of possible planetary inclinations into abundance estimates (Schneider~1994; Kostov et al.~2014; Armstrong et al.~2014, Martin \& Triaud 2014). 

In terms of stellar astrophysics, the transiting CBPs provide excellent measurements of the sizes and masses of their stellar hosts, and can notably contribute towards addressing a known tension between the predicted and observed characteristics of low-mass stars, where the stellar models predict smaller (and hotter) stars than observed (Torres et al.~2010, Boyajian et al.~2012, but also see Tal-Or et al.~2013). Each additional CBP discovery sheds new light on the still-uncertain mechanism for the formation of close binary systems (Tohline 2002). For example, the lack of CBPs around a short-period binary star (period less than $\sim7$ days) lends additional support for a commonly favored binary formation scenario of a distant stellar companion driving tidal friction and Kozai-Lidov circularization of the initially wide host binary star towards its current close configuration (Kozai~1962; Lidov~1962; Mazeh \& Shaham~1979, Fabrycky \& Tremaine~2007, Martin et al.~2015, Mu\~{n}oz \& Lai~2015, Hamers et al.~2015). 

Here we present the discovery of the Jupiter-size transiting CBP \kicb that orbits its 11.2588-day host EB every $\sim$1,100 days -- the longest-period transiting CBP at the time of writing. The planet completed a single revolution around its binary host during {\it Kepler's} data collection and was at inferior conjunction only twice -- at the very beginning of the mission (Quarter 1) and again at the end of Quarter 13. The planet transited the secondary star during the same conjunction and both the primary and secondary stars during the second conjunction. 

The first transit of the CBP \kicb was identified and reported in Welsh et al.~(2012); the target was subsequently scheduled for short cadence observations as a transiting CBP candidate (Quarters 13 through 17). At the time, however, this single event was not sufficient to rule out contamination from a background star or confirm the nature of the signal as a transit of a CBP. As {\it Kepler} continued observing, the CBP produced a second transit -- with duration and depth notably different from the first transit -- suggesting a planet on either $\sim$550-days or $\sim$1,100-days orbit. The degeneracy stemmed from a gap in the data where a planet on the former orbit could have transited (e.g.,~Welsh et al.~2014, Armstrong et al.~2014). After careful visual inspection of the \kepler light-curve, we discovered another transit, heavily blended with a primary stellar eclipse a few days before the second transit across the secondary star. As discussed below, the detection of this blended transit allowed us to constrain the period and pin down the orbital configuration of the CBP -- both analytically and numerically. 

This paper is organized as follows. We describe our analysis of the \kepler data (Section \ref{sec:kepler}) and present our photometric and spectroscopic observations of the target (Section \ref{sec:followup}). Section \ref{sec:pre_pd} details our analytical and photometric-dynamical characterization of the CB system, and outlines the orbital dynamics and long-term stability of the planet. We summarize and discuss our results in Section \ref{sec:results} and draw conclusions in Section \ref{sec:conclusions}. 

\section{{\em Kepler} Data}
\label{sec:kepler}

\begin{figure}
\centering
\plotone{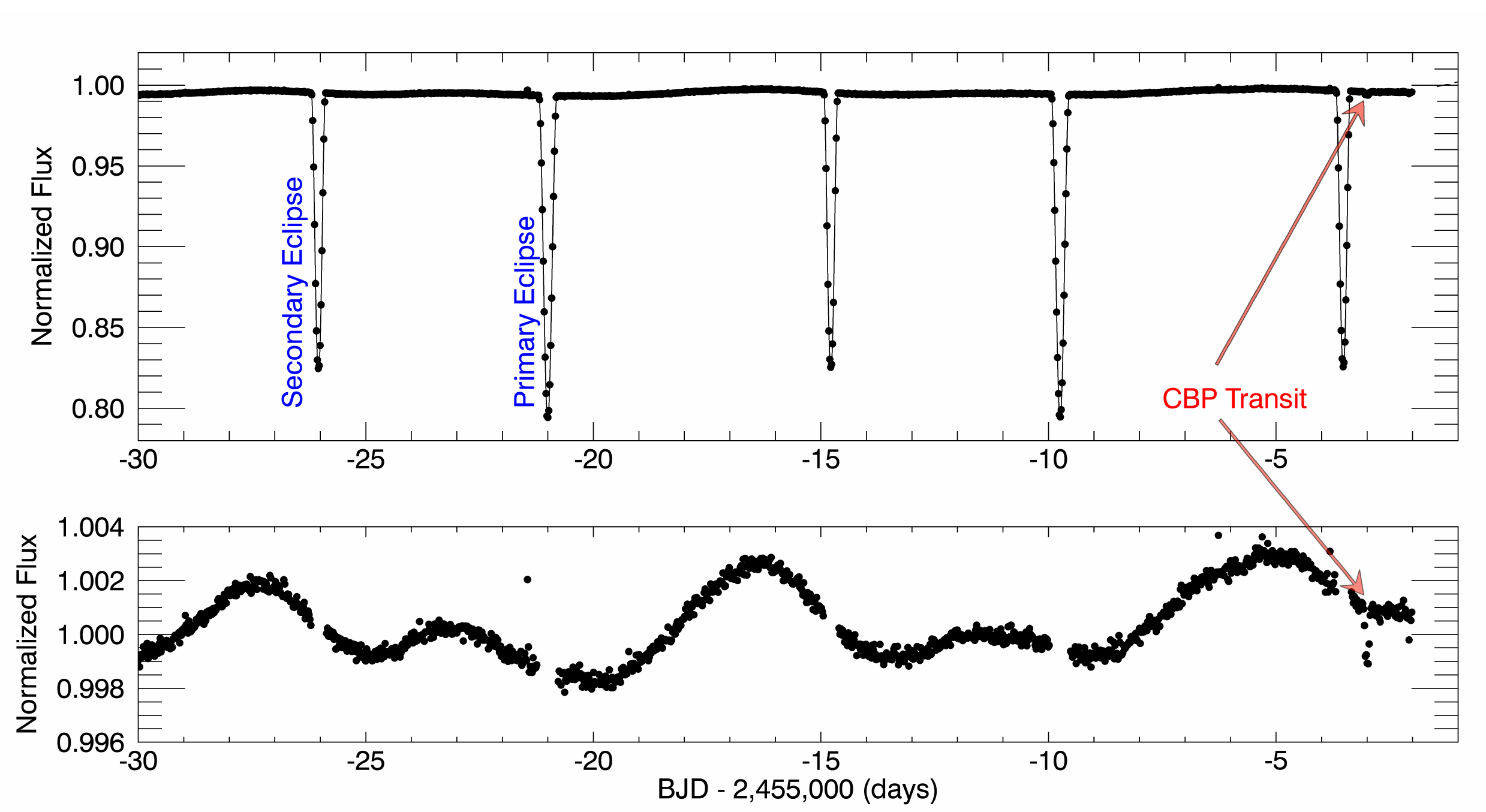}
\caption{Upper panel: A representative section of the raw (SAPFLUX), long-cadence light-curve of \kic (black symbols) exhibiting two primary, three secondary stellar eclipses, and the first transit of the CBP. Lower panel: Same, but with the stellar eclipses removed and zoomed-in to show the $\sim11-$days out-of eclipse modulation. This represents the end of Quarter 1, which is followed by a few days long data gap. Note the differences in scale between the two panels.
\label{fig:sap_LC}}
\end{figure}

\kic is listed in the NExScI Exoplanet Archive as a 11.2588-day period eclipsing binary with a {\it Kepler} magnitude of 13.545. It has an estimated effective temperature of 6217K, surface gravity log{\it g} of 4.052, metallicity of -0.78, and primary radius of 1.464 \Rsun. The target is classified as a detached eclipsing binary in the \kepler EB Catalog, with a morphology parameter $c=0.21$ (Pr{\v s}a et al.~2011, Slawson et al.~2011, Matijevi\^c et al.~2012, Kirk et al.~2015). The light-curve of \kic exhibits well-defined primary and secondary stellar eclipses with depths of $\sim20\%$ and $\sim17\%$ respectively, separated by 0.5526 in phase (see \kepler EB Catalog). A section of the raw (SAPFLUX) {\it Kepler} light-curve of the target, containing the prominent stellar eclipses and the first CBP transit, is shown in the upper panel of Figure~\ref{fig:sap_LC}.

\subsection{Stellar Eclipses}
\label{eclipses}

\begin{figure}
\centering
\plotone{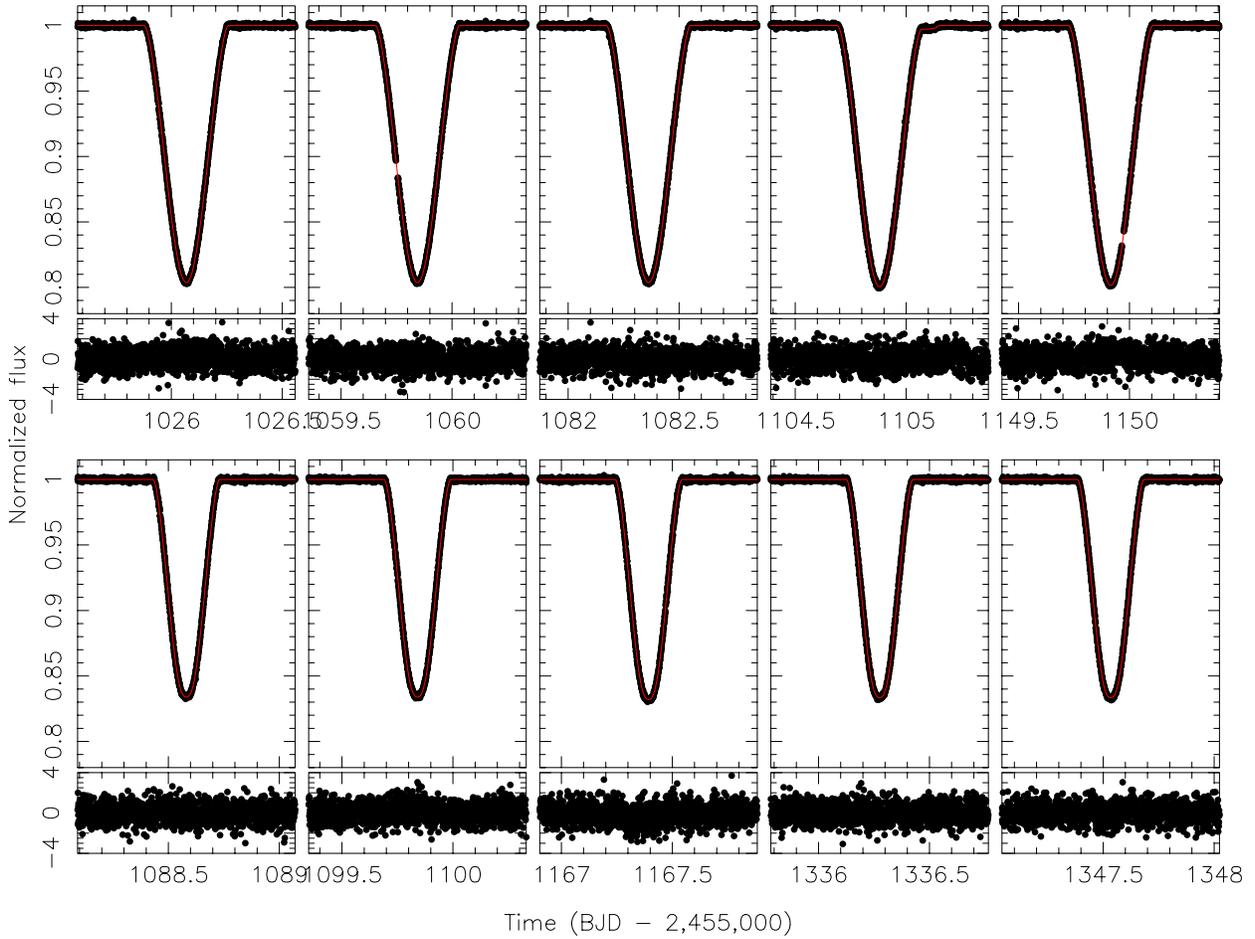}
\caption{Representative sample of primary (upper row) and secondary (lower row) stellar eclipses in short-cadence data along with our best-fit model (red) and the respective residuals. The black symbols represent normalized SC flux as a function of time (BJD - 2,455,000). A blended transit of the CBP can be seen during primary stellar eclipse near day 1105 (upper row, fourth panel from the left; the model includes the planetary transit).
\label{fig:eclipses}}
\end{figure}

The information contained in the light-curve of \kic allowed us to measure the orbital period of the EB and obtain the timing of the stellar eclipse centers ($T_{\rm prim}$, $T_{\rm sec}$), the flux and radius ratio between the two stars ($F_{\rm B}/F_{\rm A}$ and $R_{\rm B}/R_{\rm A}$), the inclination of the binary ($i_{\rm bin}$), and the normalized stellar semi-major axes ($R_{\rm A}/a_{\rm bin}$, $R_{\rm B}/a_{\rm bin}$) as follows.\footnote{Throughout this paper we refer to the binary with a subscript ``bin'', to the primary and secondary stars with subscripts ``A'' and ``B'' respectively, and to the CBP with a subscript ``p''.} First, we extracted sub-sections of the light-curve containing the stellar eclipses and the planetary transits. We only kept data points with quality flags less than 16 (see \kepler user manual) -- the rest were removed prior to our analysis. Next, we clipped out each eclipse, fit a 5th-order Legendre polynomial to the out-of-eclipse section only, then restored the eclipse and normalize to unity. Finally, we modeled the detrended, normalized and phase-folded light-curve using the ELC code (Orosz \& Hauschildt~2000, Welsh et al.~2015). A representative sample of short-cadence (SC) primary and secondary stellar eclipses, along with our best-fit model and the respective residuals, are shown in Figure~\ref{fig:eclipses}.

To measure the individual mid-eclipse times, we first created an eclipse template by fitting a Mandel \& Agol~(2002) model to the phase-folded light-curve for both the primary and secondary stellar eclipses. We carefully chose five primary and secondary eclipses (see Figure~\ref{fig:eclipses}) where the contamination from spot activity -- discussed below -- is minimal. Next, we slid the template across the light-curve, iteratively fitting it to each eclipse by adjusting only the center time of the template while kepting the EB period constant. The results of the SC and LC data analysis were merged with preference given to the SC data. 

We further used the measured primary and secondary eclipse times to calculate eclipse time variations (ETVs). These are defined in terms of the deviations of the center times of each eclipse from a linear ephemeris fit through all primary and all secondary mid-eclipse times, respectively (for a common binary period). 

\begin{figure}
\centering
\plotone{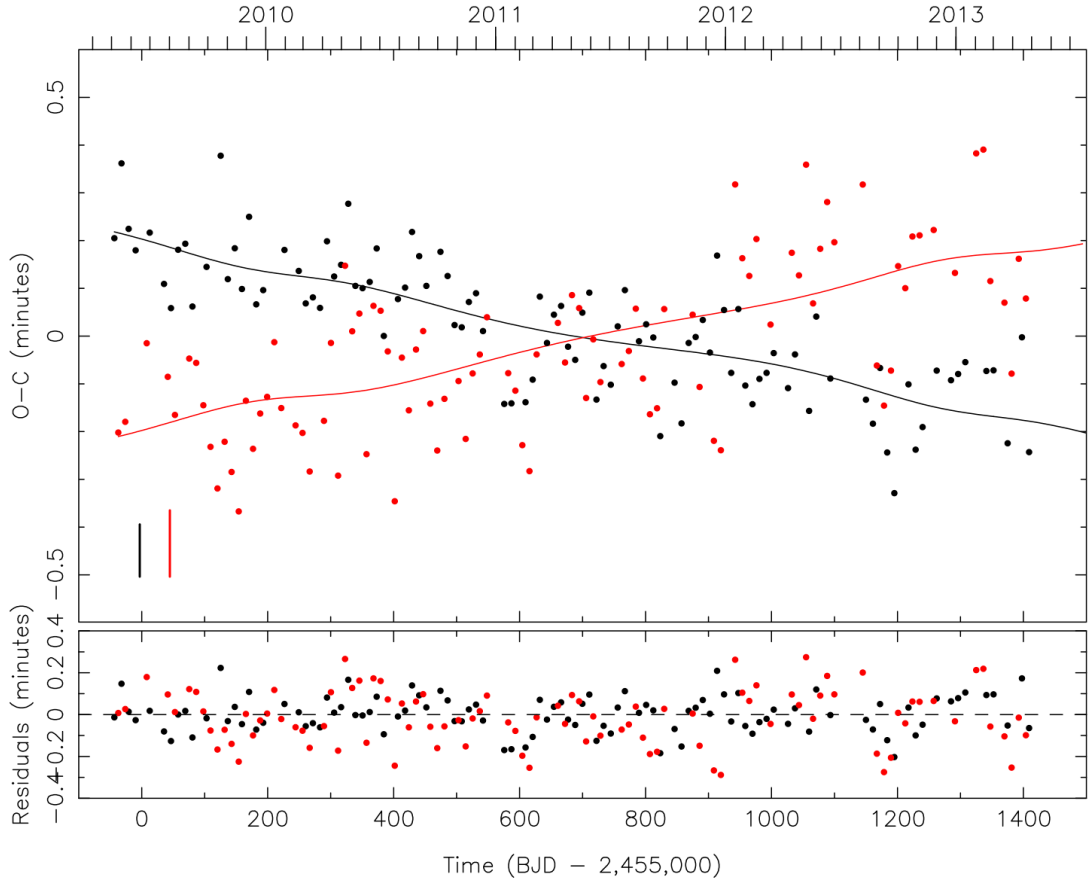}
\caption{Upper panel: Measured ``Common Period Observed minus Calculated'' (CPOC, or O-C for short) for the primary (black symbols) and secondary (red symbols) stellar eclipses; the respective lines indicate the best-fit photodynamical model. The divergent nature of the CPOCs constrains the mass of the CBP. Lower panel: CPOC residuals based on the photodynamical model. The respective average error bars are shown in the lower left of the upper panel.
\label{fig:etvs}}
\end{figure}

The respective primary and secondary ``Common Period Observed minus Calculated'' (CPOC, or O-C for short) measurements are shown in Figure~\ref{fig:etvs}. The $1\sigma$ uncertainty is $\sim0.1$~min for the measured primary eclipses, and $\sim0.14$~min for the measured secondary eclipses. As seen from the figure, the divergence in the CPOC is significantly larger than the uncertainties. 

The measured Common Period Observed minus Calculated (CPOC) are a key ingredient in estimating the mass of the CBP. As in the case of \keplerp-16b, \keplerp-34b and \keplerp-35b (Doyle et al.~2011, Welsh et al.~2012), the gravitational perturbation of \kicb imprints a detectable signature on the measured ETVs of the host EB -- indicated by the divergent primary (black symbols) and secondary (red symbols) CPOCs shown in Figure~\ref{fig:etvs}. We note that there is no detectable ``chopping'' in the ETVs (Deck \& Agol 2015) and, given that the CBP completed a single revolution during the observations, interpreting its effect on the ETVs is not trivial. However, the divergence in the CPOC residuals cannot be fully explained by a combination of general relativity (GR) correction\footnote{The GR contribution is fixed for the respective masses, period and eccentricity of the binary star} and classical tidal apsidal motion. The former is the dominant effect, with analytic $\Delta\omega_{\rm GR} = 0.00019$  deg/cycle, while the latter has an analytic contribution of $\Delta\omega_{\rm tidal} = 0.00003$ deg/cycle (using the best-fit apsidal constants of $k_{2}(\rm A) = 0.00249$ for the slightly evolved primary and $k_{2}(\rm B) =  0.02978$ for the secondary respectively). 

The total analytic precession rate, $\Delta\omega_{\rm analytic} = \Delta\omega_{\rm GR} + \Delta\omega_{\rm tidal} = 0.00022$ deg/cycle, is $\approx9.5\%$ smaller than the numeric rate of $\Delta\omega_{\rm numeric} = 0.00024$ (deg/cycle), as calculated from our photodynamical model (Section 4). The difference represents the additional push from the CBP to the binary's apsidal motion and allowed us to evaluate the planet's mass. Thus the relative contributions to the apsidal precession of the binary star are $\Delta\omega_{\rm GR} = 77.4\%$, $\Delta\omega_{\rm tidal} = 13.9\%$, and $\Delta\omega_{\rm CBP} = 8.7\%$. The difference between the analytic and numerically determined precession rates is significant -- for a test-mass planet the latter agrees with the former to within $0.5\%$.

\begin{figure}
\centering
\epsscale{0.7}
\plotone{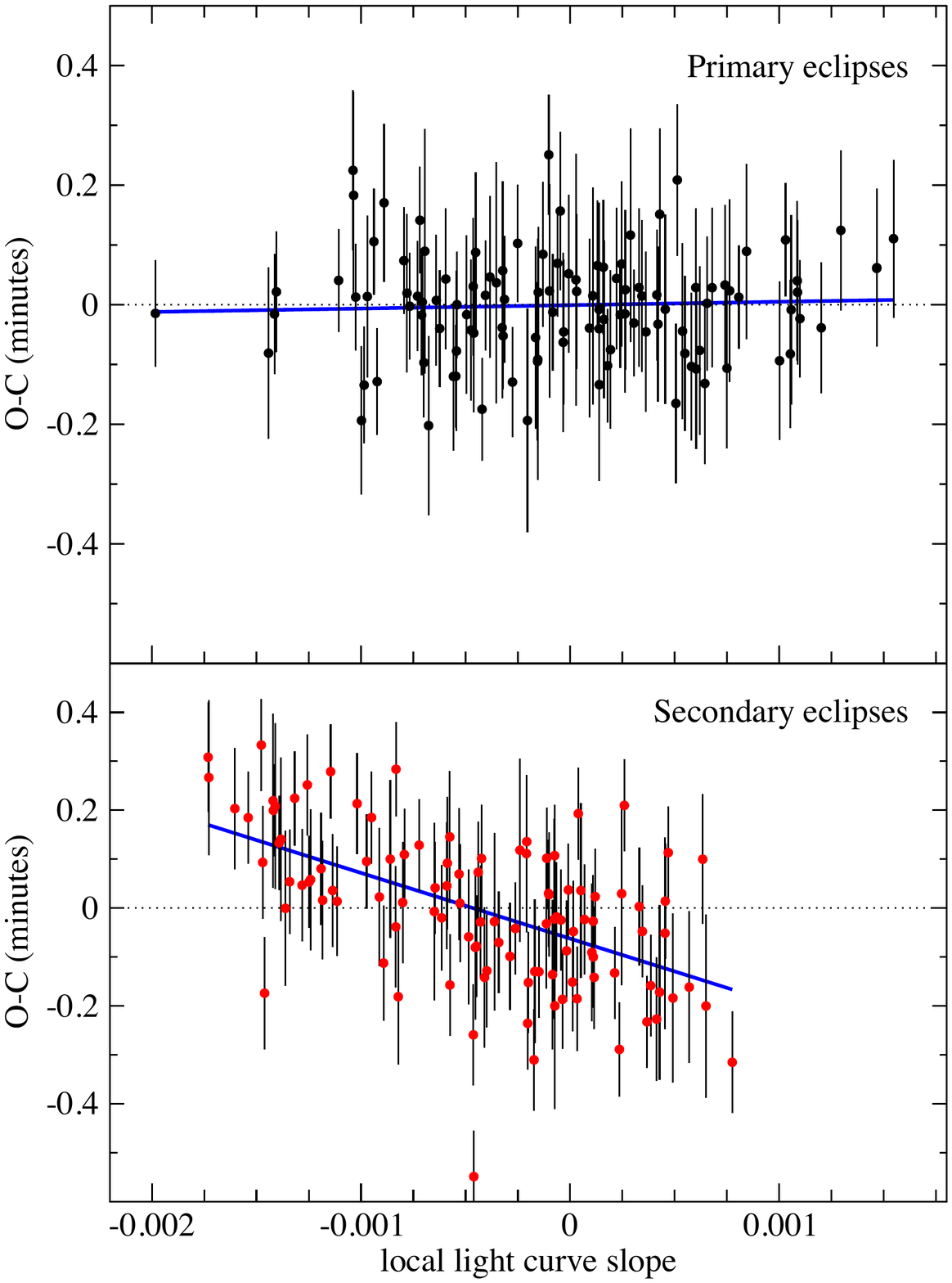}
\caption{Measured local slope (see text for details) of the light-curve during primary (upper panel) and secondary (lower panel) stellar eclipses as a function of the respective Observed minus Calculated (``O-C'') eclipse times. The anti-correlation seen for the secondary eclipses indicates that the modulations seen in the light-curve are caused by the rotation of the secondary star.
\label{fig:spot_slope}}
\end{figure}

\begin{figure}
\centering
\epsscale{1.1}
\plotone{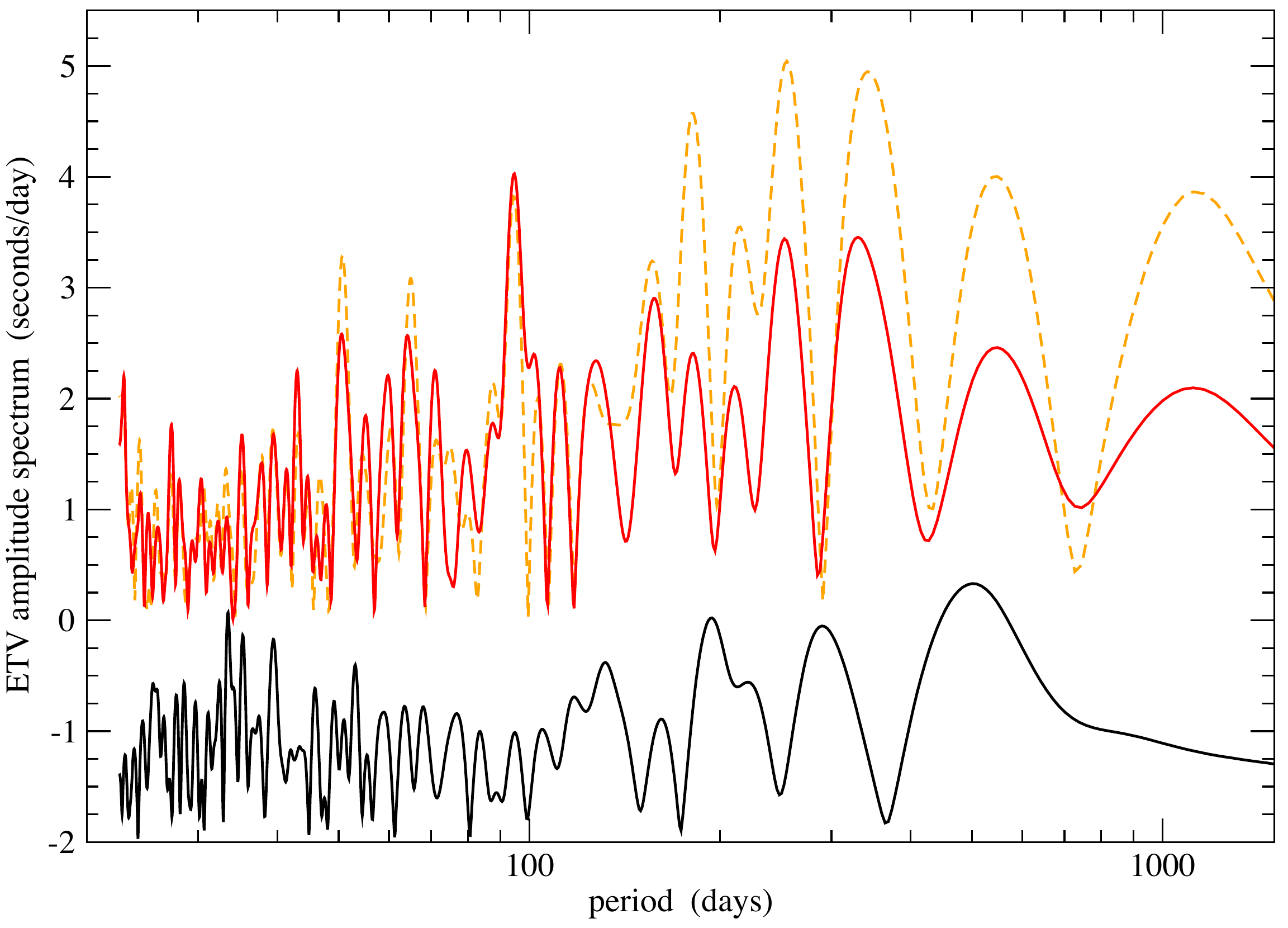}
\caption{Power spectrum of the primary (black line) and secondary (solid red and dashed orange lines) O-C eclipse times (based on a linear ephemeris). The primary power spectrum is offset vertically for viewing purposes. The dashed orange and solid red lines represent the secondary ETV power spectrum before and after correcting for the anti-correlation between the local slope of the lightcurve and the measured secondary eclipse times (see Figure~\ref{fig:spot_slope}). There are no statistically significant peaks in either the primary or corrected secondary ETVs.
\label{fig:ls_etvs}}
\end{figure}

Following the method of Welsh et al.~(2015), we also analyzed the effect of star spots on the measured eclipse times by comparing the local slope of the light-curve outside the stellar eclipses to the measured ETVs (see also Orosz et al.~2012; Mazeh, Holczer, Shporer~2015; Holczer, et al.~2015). While there is no apparent correlation for the primary eclipse ETVs (Figure~\ref{fig:spot_slope}, left panel), we found a clear, negative correlation for the secondary eclipses (Figure~\ref{fig:spot_slope}, right panel), indicating that the observed light-curve modulations are due to the rotationally-modulated signature of star spots moving across the disk of the secondary star. The negative correlation also indicates that the spin and orbital axes of the secondary star are well aligned (Mazeh et al.~2015, Holczer et al.~2015). We have corrected the secondary eclipse times for this anti-correlation. We note that as the light from the secondary is diluted by the primary star, its intrinsic photometric modulations are larger than the observed $0.1-0.3\%$, indicating a rather active secondary star. The power spectrum of the O-C ETVs (in terms of measured primary and secondary eclipse times minus linear ephemeris) are shown in Figure~\ref{fig:ls_etvs}. There are no statistically significant features in the power spectrum.

The exquisite quality of the {\it Kepler} data also allowed for precise measurement of the photometric centroid position of the target. Apparent shifts in this position indicate either a contamination from nearby sources (if the centroid shifts away from the target during an eclipse or transit), or that the source of the studied signal is not the \kepler target itself (if the shift is towards the target during an eclipse or transit, see Conroy et al.~2014). As expected, \kic exhibits a clear photocenter shift away from the target and toward the nearby star during the eclipses, indicating that the eclipses are indeed coming from the target stars, and that some light contamination from a nearby star is present in the {\it Kepler} aperture. We discuss this in more details in Section \ref{sec:phot_followup}.

\subsection{Stellar Rotation}
\label{sec:rot}

\begin{figure}
\centering
\epsscale{1.1}
\plotone{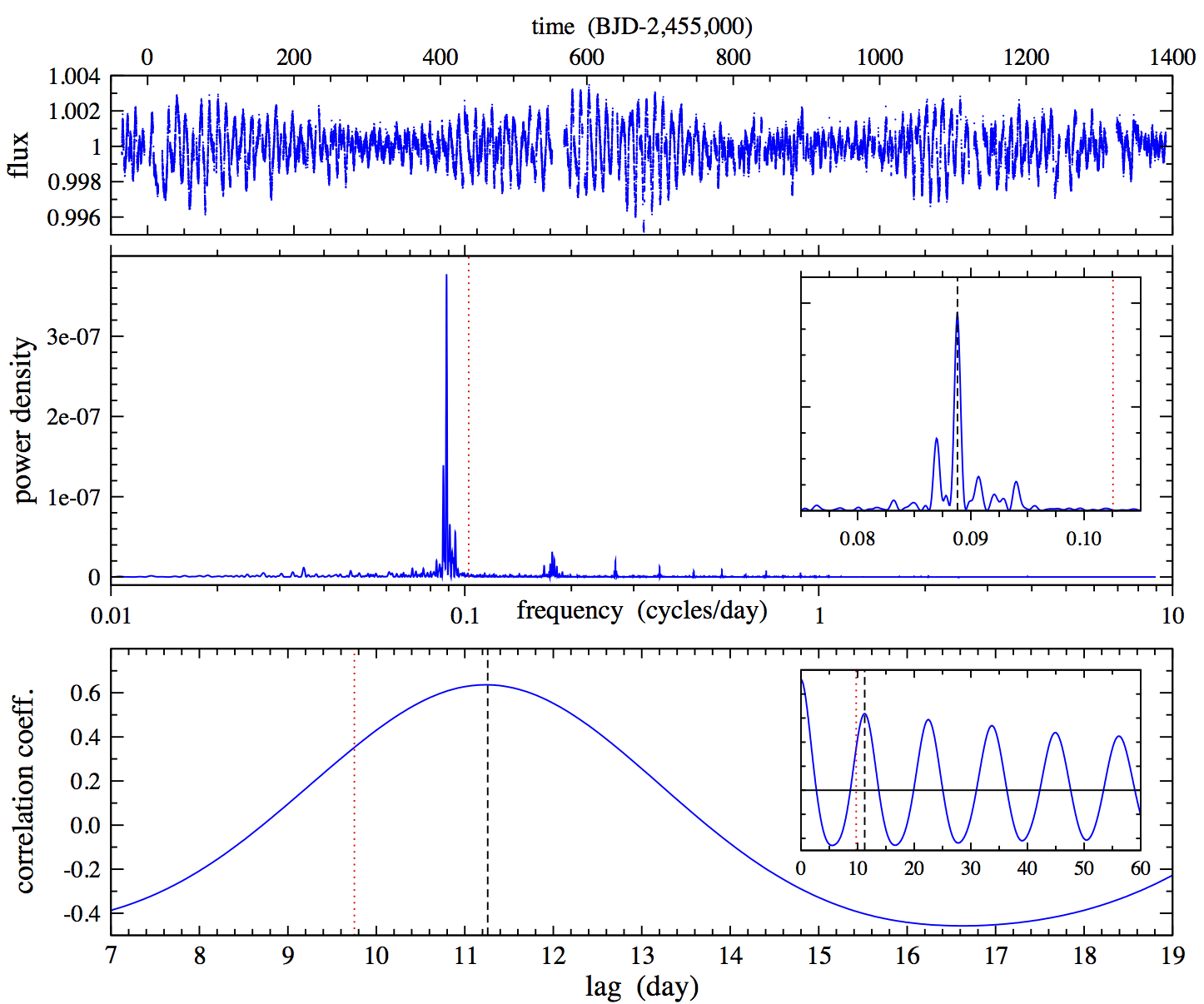}
\caption{Upper panel: Mildly detrended, normalized raw (SAPFLUX) light-curve for the Q1-16 \kepler data; middle panel: L-S periodogram of the out-of-eclipse regions of the light-curve, revealing a clear peak near 11 days; lower panel: ACF of the light-curve. The inset in the middle panel is zoomed out to better represent the long-term stability of the ACF modulation. around the L-S spike. The vertical lines in the middle and lower panels indicate the binary period (black dashed line) -- overlapping with both the L-S and the ACF peak -- and the pseudosynchronous period (red dotted line). The L-S and ACF periods are consistent with each other with the binary period, and differ significantly from pseudosynchronicity.\label{fig:ACF}}
\end{figure}

The out-of-eclipse sections of the light-curve are dominated by quasi-periodic flux modulations with an amplitude of $0.1-0.3\%$ (lower panel, Figure~\ref{fig:sap_LC}). To measure the period of these modulations we performed both a Lomb-Scargle (L-S) and an autocorrelation function (ACF) analysis of the light-curve. For the latter method, based on measuring the time lags of spot-induced ACF peaks, we followed the prescription of McQuillan et al.~(2013, 2014). Both methods show clear periodic modulations (see Figure~\ref{fig:sap_LC} and Figure~\ref{fig:ACF}), with $P_{\rm rot} = 11.23\pm0.01$~days -- very close to the orbital period of the binary. We examined the light-curve by eye and confirmed that this is indeed the true period. 

To measure the period, we first removed the stellar eclipses from the light curve using a mask that was 1.5 times the duration of the primary eclipse (0.220 days) and centered on each eclipse time. After the eclipses were removed, a 7th order polynomial was used to moderately detrend each Quarter. In Figure~\ref{fig:ACF} we show the mildly detrended SAPFLUX light-curve covering Q1-16 \kepler data (upper panel), the power spectrum as a function of the logarithmic frequency (middle panel) and the ACF of the light-curve (lower panel). Each Quarter shown in the upper panel had a 7th order polynomial fit divided out to normalize the light-curve, and we also clipped out the monthly data-download glitches by hand, and any data with Data Quality Flag $>$8. The inset in the lower panel zooms out on to better show the long-term stability of the ACF; the vertical black dashed lines in the middle and lower panels represent the best fit orbital period of the binary star, and the red dotted lines represent the expected rotation period if the system were in pseudosynchronous rotation. The periods as derived from the ACF and from L-S agree exactly. 

The near-equality between the orbital and the rotation period raises the question whether the stars could be in pseudosynchronous pseudo-equilibrium. If this is the case, then with $P_{\rm bin} = 11.258818$ days and $e_{\rm bin}=0.16$, and based on Hut's formula (Hut 1981, 1982), the expected pseudosynchronous period is $P_{\rm pseudo, rot} = 9.75$ days. Such spin-orbit synchronization should have been reached within a Gyr\footnote{Orbital circularization takes orders of magnitude longer and is not expected.}. Thus the secondary star is not rotating pseudosynchronously. This is clearly illustrated in Figure~\ref{fig:ACF}, where the red dotted line in the inset in the lower panel represents the pseudosynhronous rotation period.

Our measurement of the near-equality between $P_{\rm rot}$ and $P_{\rm bin}$ indicates that the rotation of both stars is synchronous with the binary period (due to tidal interaction with the binary orbit). Thus the secondary (G-type) star appears to be tidally spun up since its rotation period is faster than expected for its spectral type and age (discussed in more detail in Section 5), and has driven the large stellar activity, as seen by the large amplitude star spots. The primary star does not appear to be active. The spin-up of the primary (an F-type star) should not be significant since F-stars naturally rotate faster, and are quieter than G-stars (assuming the same age). As a result, the primary could seem quiet compared to the secondary -- but this can change as the primary star evolves. As seen from Figure~\ref{fig:ACF}, the starspot modulation is indeed at the binary orbital period (black dashed line, inset in lower panel). 

There is reasonable evidence that stars evolving off the main sequence look quieter than main sequence stars (at least for a while); stars with shallower convection zones look less variable at a given rotation rate (Bastien et al.~2014). The convective zone of the primary star is probably too thin for significant spot generation at that rotation period. As we show in Section \ref{sec:results}, the secondary star has mass and effective temperature very similar to the Sun, so it should be generating spots at about the same rate as the Sun would have done when it was at the age of NGC 6811 -- where early G stars have rotation periods of 10-12 days (Meibom et al 2011). 

Based on the photodynamically-calculated stellar radii $R_{\rm A}$ and $R_{\rm B}$ (Section \ref{sec:results}), and on the measured $P_{\rm rot}$, if the two stars are indeed synchronized then their rotational velocities should be $V_{\rm rot,A}\sin{i_{\rm A}} = 8.04$ km/s and $V_{\rm rot,B}\sin{i_{\rm B}} = 4.35$ km/s. If, on the contrary, the two stars are rotating pseudosynchronously, their respective velocities should be $ V_{\rm rot,A}\sin{i_{\rm A}} =  9.25$ km/s and $V_{\rm rot,B}\sin{i_{\rm B}} =  5.00$ km/s. 

The spectroscopically-measured rotational velocities (Section \ref{sec:spec_followup}) are $V_{\rm rot,A}\sin{i_{\rm A}} = 8.4\pm0.5$ km/s and $V_{\rm rot,B}\sin{i_{\rm B}} = 5.1\pm1.0$ km/s respectively -- assuming 5.5 km/s macroturbulence for the primary star (Doyle et al.~2014), and 3.98 km/s macroturbulence for Solar-type stars (Gray~1984) as appropriate for the secondary. Given the uncertainty on both the measurements and the assumed macroturbulence, the measured rotational velocities are not inconsistent with synchronization. 

Combined with the measured rotation period, and assuming spin-orbit synchronization of the binary, the measured broadening of the spectral lines constrain the stellar radii: 

\begin{equation}
R_{\rm B} = \frac{P_{\rm rot}V_{\rm rot,B}\sin{i_{\rm B}}}{2\pi\phi}
\label{eq:R_rot}
\end{equation}

{\noindent where $\phi$ accounts for differential rotation and is a factor of order unity.} Assuming $\phi=1$, $R_{\rm B}=1.1\pm0.2~\Rsun$ and $R_{\rm A}=1.85~R_{\rm B}~(\rm see~Table~2) = 2.08\pm0.37~\Rsun$.

\section{Spectroscopic and Photometric follow-up observations}
\label{sec:followup}

To complement the {\it Kepler} data, and better characterize the \kic system, we obtained comprehensive spectroscopic and photometric follow-up observations. Here we describe the radial velocity measurements we obtained to constrain the spectroscopic orbit of the binary and calculate the stellar masses, its orbital semi-major axis, eccentricity and argument of periastron. We also present our spectroscopic analysis constraining the effective temperature, metallicity and surface gravity of the two stars, our direct-imaging observations aimed at estimating flux contamination due to unresolved background sources, and our ground-based observations of stellar eclipses to extend the accessible time baseline past the end of the original \kepler mission. 

\subsection{Spectroscopic follow-up and Radial Velocities}
\label{sec:spec_followup}

We monitored \kic spectroscopically with several instruments in order to measure the radial velocities of the two components of the EB.  Observations were collected with the Tillinghast Reflector Echelle Spectrograph \citep[TRES;][]{furesz08} on the 1.5-m telescope at the Fred L.\ Whipple Observatory, the Tull Coude Spectrograph (Tull et al.~1995) on the McDonald Observatory 2.7-m Harlan J.\ Smith Telescope, the High Resolution Echelle Spectrometer (HIRES; Vogt et al.~1994) on the 10-m Keck\,I telescope at the W.\ M.\ Keck Observatory, and the Hamilton Echelle Spectrometer (HamSpec; Vogt 1987) on the Lick Observatory 3-m Shane telescope.

A total of 14 observations were obtained with TRES (eight in 2011, five in 2012, and one in 2013). They span the wavelength range from about 390 to 900~nm at a resolving power of $R \approx 44,000$. We extracted the spectra following the procedures outlined by Buchhave et al.\ (2010). Seven observations were gathered with the Tull Coude Spectrograph in 2011, each consisting of three exposures of 1200 seconds. This instrument covers the wavelength range 380--1000~nm at a resolving power of $R \approx 60,000$. The data were reduced and extracteded with the instrument pipeline. Two observations were obtained with HIRES also in 2011, covering the range 300-1000~nm at a resolving power of $R \approx 60,000$ at 550~nm. We used the C2 decker for sky subtraction, giving a sky-projected area for the slit of $0\farcs87 \times 14\farcs0$. Th-Ar lamp exposures were used for wavelength calibration, and the spectra were extracteded with the pipeline used for planet search programs at that facility. Finally, two spectra were collected with HamSpec in 2011, with a wavelength coverage of 385--955~nm and a resolving power of $R \approx 60,000$.

All of these spectra are double-lined.\footnote{We note that Kolbl et al.~(2015) detected the spectrum of the secondary star as well, and obtained its effective temperature ($T_{\rm eff,B} \sim 5900$~K) and flux ratio ($F_{\rm B}/F_{\rm A} = 0.22$) -- fully consistent with our analysis.}  RVs for both binary components from the TRES spectra were derived using the two-dimensional cross-correlation technique TODCOR \citep{zucker:94}, with templates taken from a library of synthetic spectra generated from model atmospheres by R.\ L.\ Kurucz \citep[see][]{nord94, latham02}. These templates were calculated by John Laird, based on a line list compiled by Jon Morse.  The synthetic spectra cover 30~nm centered near 519~nm, though we only used the central 10~nm, corresponding to the TRES echelle order centered on the gravity-sensitive \ion{Mg}{1}\,b triplet.  Template parameters (effective temperature, surface gravity, metallicity, and rotational broadening) were selected as described in the next section.

To measure the RVs from the other three instruments, we used the ``broadening function'' (BF) technique (e.g.,~Schwamb et al.~2013), in which the Doppler shift can be obtained from the centroid of the peak corresponding to each component in the broadening function, and the rotational broadening is measured from the peak's width. This method requires a high-resolution template spectrum of a slowly rotating star, for which we used the RV standard star HD~182488 (a G8V star with a RV of +21.508~\kms; see Welsh et al.~2015). All HJDs in the Coordinated Universal Time (UTC) frame were converted to BJDs in the Terrestrial Time (TT) frame using the software tools by Eastman et al.~(2010). It was also necessary to adjust the RV zero points to match that of TRES by +0.66~\kms\ for McDonald, $-0.16$~\kms\ for Lick, and by +0.28~\kms\ for Keck. We report all radial velocity measurements in Table~\ref{tab: tab_RV}.

\subsection{Spectroscopic Parameters}
\label{sec:todcor}

In order to derive the spectroscopic parameters ($T_{\rm eff}$, $\log g$, [m/H], $v \sin i$) of the components of the \kic binary, both for obtaining the final radial velocities and also for later use in comparing the physical properties of the stars with stellar evolution models, we made use of TODCOR as a convenient tool to find the best match between our synthetic spectra and the observations. Weak spectra or blended lines can prevent accurate classifications, so we included in this analysis only the 11 strongest TRES spectra (${\rm S/N} > 20$), and note that all of these spectra have a velocity separation greater than 30~\kms\ between the two stars.

We performed an analysis similar to the one used to characterize the stars of the CBP-hosting double-lined binaries \keplerp-34 and \keplerp-35 (Welsh et al.~2012), but given slight differences in the analysis that are required by the characteristics of this system, we provide further details here. We began by cross-correlating the TRES spectra against a (five-dimensional) grid of synthetic composite spectra that we described in the previous section. The grid we used for \kic contains every combination of stellar parameters in the ranges $T_{\rm eff,A} = [4750, 7500]$, $T_{\rm eff,B} = [4250, 7250]$, $\log g_{\rm A} = [3.0, 5.0]$, $\log g_{\rm B} = [3.5, 5.0]$, and [m/H]$ = [-1.0, +0.5]$, with grid spacings of 250~K in $T_{\rm eff}$, and 0.5~dex in $\log g$ and [m/H] (12,480 total grid points).\footnote{We ran a separate TODCOR grid solely to determine the $v \sin i$ values, which we left fixed in the larger grid. This is justified because the magnitude of the covariance between $v \sin i$ and the other parameters is small. This simplification reduces computation time by almost two orders of magnitude.} At each step in the grid, TODCOR was run in order to determine the RVs of the two stars and the light ratio that produces the best-fit set of 11 synthetic composite spectra, and we saved the resulting mean correlation peak height from these 11 correlations. Finally, we interpolated along the grid surface defined by these peak heights to arrive at the best-fit combination of stellar parameters.

This analysis would normally be limited by the degeneracy between spectroscopic parameters (i.e., a nearly equally good fit can be obtained by slightly increasing or decreasing $T_{\rm eff}$, $\log g$, and [m/H] in tandem), but the photodynamical model partially breaks this degeneracy by providing precise, independently determined surface gravities. We interpolated to these values in our analysis and were left with a more manageable $T_{\rm eff}$-[m/H] degeneracy. In principle one could use temperatures estimated from standard photometry to help constrain the solution and overcome the $T_{\rm eff}$-[m/H] correlation, but the binary nature of the object (both stars contributing significant light) and uncertainties in the reddening make this difficult in practice.  In the absence of such an external constraint, we computed a table of $T_{\rm eff,A}$ and $T_{\rm eff,B}$ values as a function of metallicity, and found the highest average correlation value for ${\rm [m/H]} = -0.18$, leading to temperatures of 6190 and 5760~K for the primary and secondary of \kiccomma, respectively. The average flux ratio from this best fit is $F_{\rm B}/F_{\rm A} = 0.21$ at a mean wavelength of 519~nm. To arrive at the final spectroscopic parameters, we elected to resolve the remaining degeneracy by appealing to stellar evolution models. This procedure is described below in Section~5.1, and results in slightly adjusted values of ${\rm [m/H]} = -0.14 \pm 0.05$\footnote{Note the difference from the NexSci catalog value of -0.78.} and temperatures of 6210 and 5770~K, with estimated uncertainties of 100~K.

\begin{figure}
\centering
\plotone{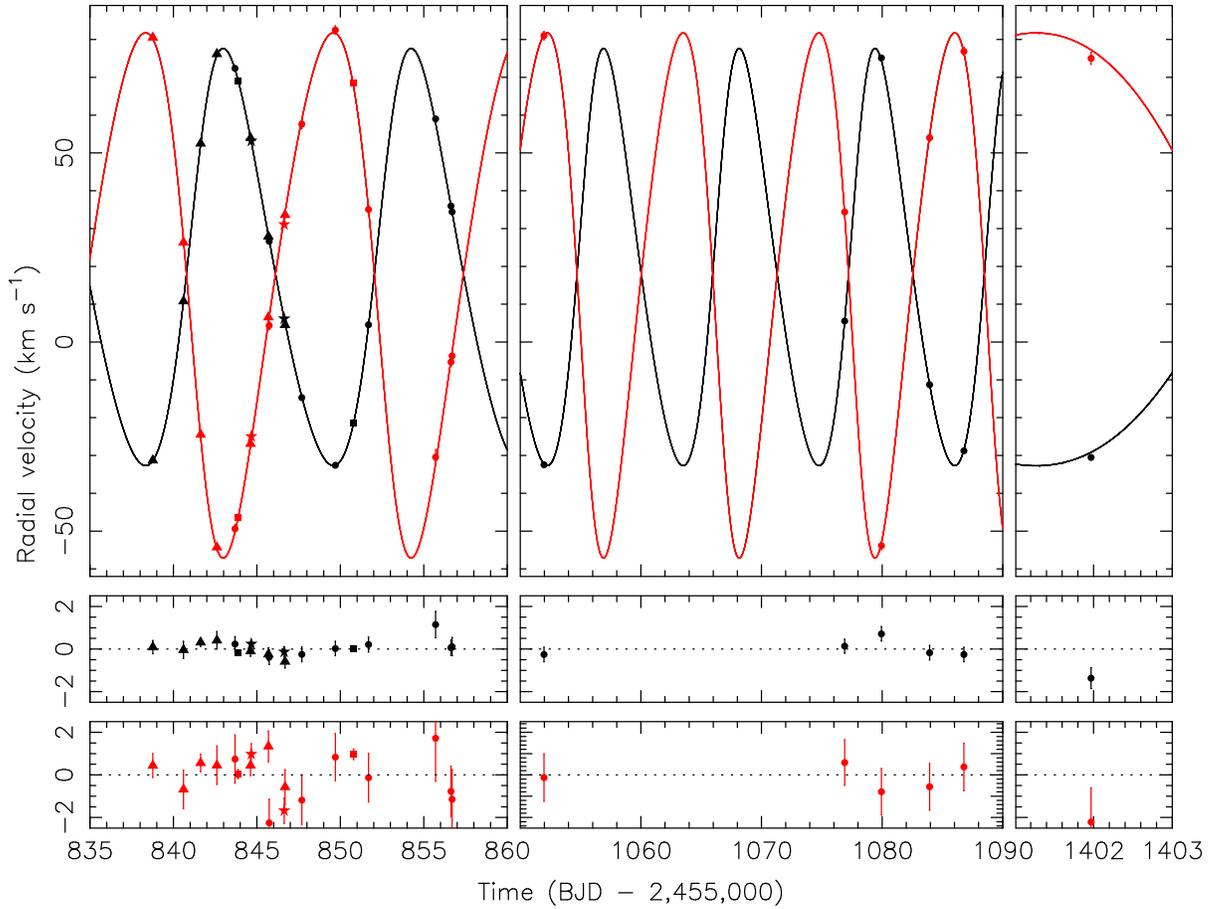}
\caption{Upper panels: Radial velocity measurements for the primary (black symbols) and secondary (red symbols) stars of the EB \kic from the McDonald 2.7-m (triangles), Lick 3-m (stars), Keck I 10-m (squares) and the Tillinghast 1.5-m telescopes (circles), and the respective Keplerian fits (solid lines); Lower panels:  $1\sigma$ residuals between the measured RVs and their corresponding best fits.
\label{fig:rvs}}
\end{figure}

\subsection{Direct imaging follow-up}
\label{sec:phot_followup}

Due to {\it Kepler's} large pixel size (3.98\arcsec, Koch et al.~2010), it is possible for unresolved sources to be present inside the target's aperture, and to also contaminate its light-curve. A data query from MAST indicates that the \kic suffers a mean contamination of $4\pm1\%$ between the four seasons. To fully account for the effect this contamination has on the inferred sizes of the occulting objects, we performed an archival search and pursued additional photometric observations.  

A nearby star to the south of \kic is clearly resolved on UKIRT/WFCAM J-band images (Lawrence et al.~2007), with $\Delta J = 2.2$ mag and separation of $2.8\arcsec$. \kic was also observed in g-, r-, i-bands, and $H_{\alpha}$ from the INT survey (Greiss et al.~2012). The respective magnitude differences between the EB and the companion are $\Delta g = 3.19, \Delta r = 2.73, \Delta H_{\alpha} = 2.61$, and $\Delta i = 2.52$ mag with formal uncertainties below 1\%. Based on Equations 2 through 5 from Brown et al.~(2011) to convert from Sloan to $K_{\rm p}$, these correspond to magnitude and flux differences between the EB and the companion star of $\Delta K_{\rm p} = 2.73$ and $\Delta F = 8\%$. In addition, adaptive-optics observations by Dressing et al.~(2014) with MMT/ARIES  detected the companion at a separation of $2.78\arcsec$ from the target, with $\Delta K_{\rm s} = 1.84$ mag, estimated $\Delta K_{\rm p} = 2.2$ mag, and reported a position angle for the companion of $131.4\arcdeg$ (East of North).

\begin{figure}
\centering
\epsscale{0.5}
\plotone{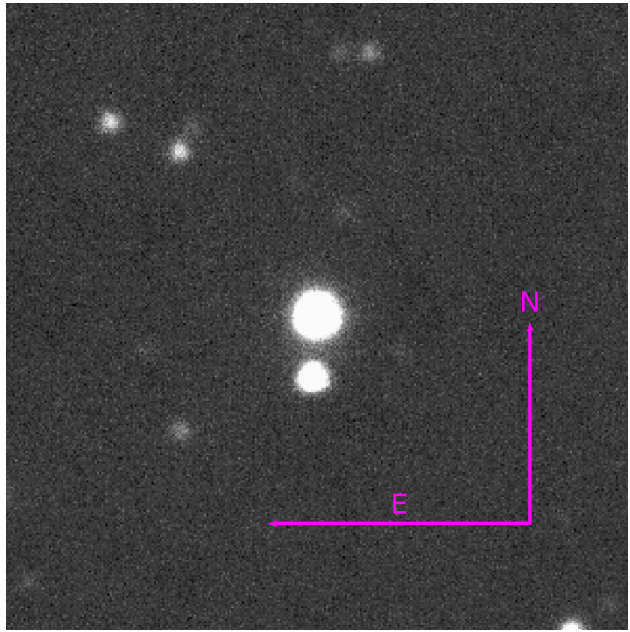}
\caption{A J-band WIYN/WHIRC image of \kic showing the nearby star to the south of the EB. The size of the box is $30\arcsec$ by $30\arcsec$. The two stars are separated by $2.89\pm0.14\arcsec$.
\label{fig:wiyn_J}}
\end{figure}

We observed the target with WIYN/WHIRC (Meixner et al.~2010) on 2013, Oct, 20 (UT), using a five-point dithering pattern, J, H and Ks filters, and 30 sec of integration time; the seeing was $0.73\arcsec$ (J-band), $0.72\arcsec$ (H-band) and $0.84\arcsec$ (Ks-band). We confirmed the presence of the companion (see Figure~\ref{fig:wiyn_J}), and obtain a magnitude difference of $\Delta J = 2.21\pm0.04$ mag, $\Delta H = 1.89\pm0.06$ mag, and $\Delta K_{\rm s} = 1.85\pm0.11$ mag respectively. Using the formalism of Howell et al.~(2012), we estimated $\Delta K_{\rm p} = 2.85$ mag if the companion is a giant and $\Delta K_{\rm p} = 2.9$ mag if it is a dwarf star. We adopt the latter -- i.e., flux contamination of $6.9\pm1.5\%$ -- for our pre-photodynamical analysis of the system. The position angle of the companion from our Ks-band WIYN/WHIRC images is $176.02\pm0.23\arcdeg$ (East of North) -- consistent with the UKIRT J-band data where the position angle of the companion is $\approx176\arcdeg$, and notably different from the results of Dressing et al.~(2014).

We evaluated the probability for the companion star to be randomly aligned on the sky with \kic using the estimates of Gilliland et al.~(2011) for the number of blended background stars within a target's aperture. At the Galactic latitude of \kic (b = 6.84\arcdeg), there is $\approx1.1\%$ chance for a random alignment between \kic and a background source of $K_{\rm p} \le 16.45$ separated by $2.8\arcsec$, suggesting that this source is likely to be a bound companion to \kic.

As mentioned in Sec. \ref{sec:kepler}, there is a noticeable photometric centroid shift in the photometric position of \kic during the stellar eclipses. To investigate this we examined the NASA Exoplanet Archive Data Validation Report (Akeson et al.~2013) for \kic. The report provides information on the location of the eclipse signal from two pixel-based methods -- the photometric centroid\footnote{Which tracks how the center of light changes as the amount of light changes, e.g.,~during eclipses.} and the pixel-response function (PRF) centroid (Bryson et al.~2013)\footnote{Which tracks the location of the eclipse source. Specifically, the centroid of the PRF difference image (the difference between the out-of transit (OoT) and in-transit images) indicates the location of the eclipse source and the PRF OoT centroid indicates the location of the target star. Differences between these centroids provide information on the offset between the eclipse source and the target star.}.

The photometric centroid of \kic has an RA offset of $0.03\arcsec$ and a Dec offset of $-0.05\arcsec$ in-eclipse. The PRF difference image centroid is offset relative to the PRF OoT centroid by RA = $-0.012\arcsec$ and Dec = $0.136\arcsec$ respectively (the offsets relative to the KIC position are RA = $-0.019\arcsec$ and Dec = $0.01\arcsec$). Both methods indicate that the measured center of light shifts away from \kic during the stellar eclipses, fully consistent with the photometric contamination from the companion star to the SE of the EB.

\subsection{Time-series photometric follow-up}
\label{sec:KELT_followup}

In order to confirm the model derived from the \kepler data and to place additional constraints on the model over a longer time baseline, we undertook additional time-series observations of \kic using the KELT follow-up network, which consists of small and mid-size telescopes used for confirming transiting planets for the KELT survey (Pepper, et al.~2007; Siverd, et al.~2012).  Based on predictions of primary eclipse times for \kiccomma, we obtained two observations of partial eclipses after the end of the \kepler primary mission.  The long durations of the primary eclipse ($>9$ hours) make it nearly impossible to completely observe from any one site, but partial eclipses can nevertheless help constrain the eclipse time.

We observed a primary eclipse simultaneously in V and i at Swarthmore College's Peter van de Kamp Observatory on UT2013-08-17. The observatory uses a 0.6 m RCOS Telescope with an Apogee U16M $4K\times4K$ CCD, giving a $26\rm \arcmin\times26\rm \arcmin$ field of view. Using $2\times2$ binning, it has 0.76$\rm \arcsec/\rm pixel$. These observations covered the second half of the eclipse, extending about 2.5 hours after egress. 

\begin{figure}
\centering
\epsscale{0.5}
\plotone{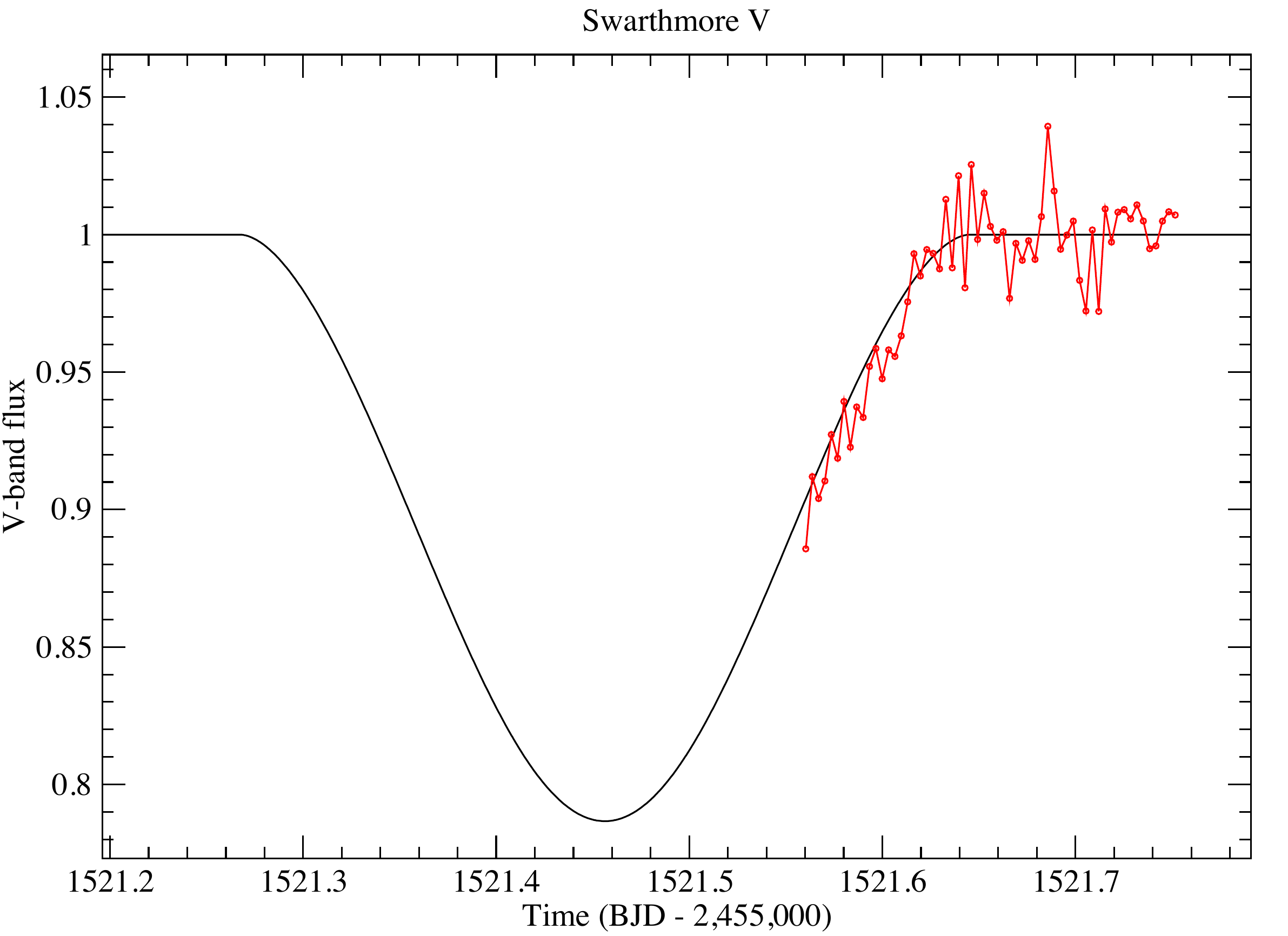}
\plotone{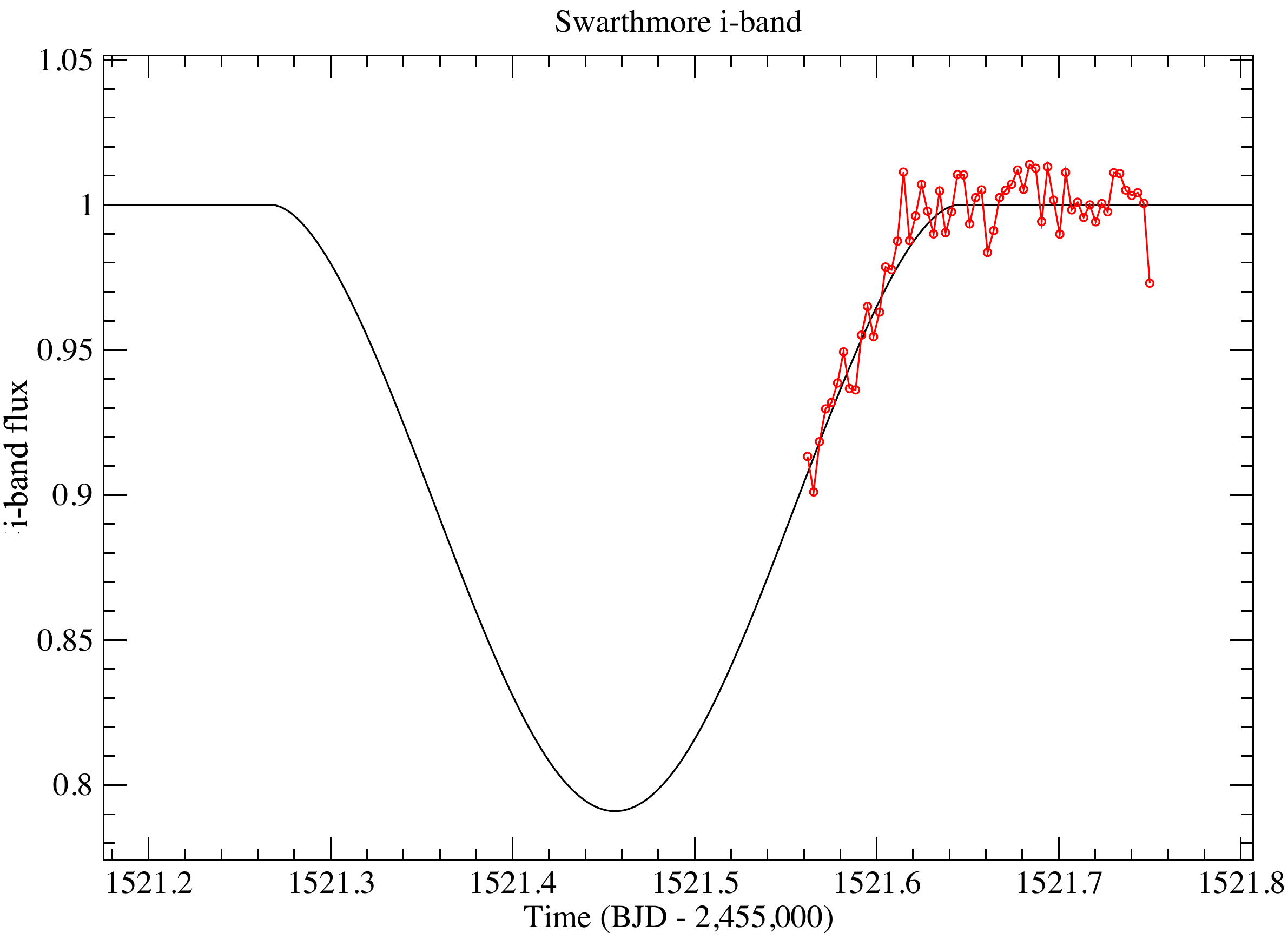}
\plotone{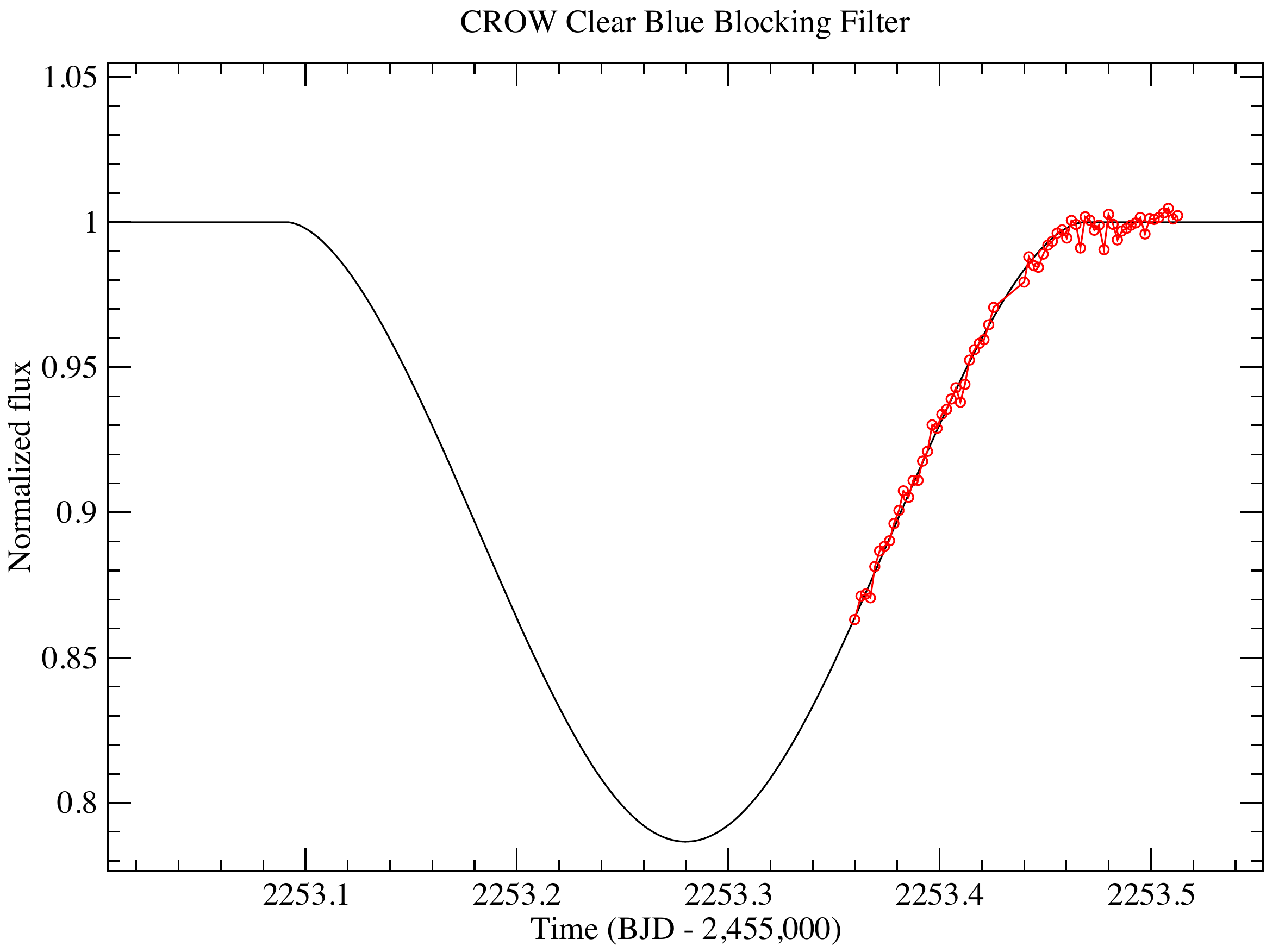}
\caption{Observed (KELT network data, red symbols) and photodynamically-predicted (solid line) primary eclipses of \kic. The upper panel is for Swarthmore V-band, the middle -- for Swarthmore i-band, and the lower --  CROW observatory CBB filter (see text for details). The data are fully consistent with the photodynamical predictions.
\label{fig:KELT}}
\end{figure}

We observed a primary eclipse of the system at Canela's Robotic Observatory (CROW) in Portugal. Observations were made using a 0.3 m LX200 telescope with a SBIG ST-8XME CCD. The FOV is $28\rm \arcmin\times19\rm \arcmin$ and 1.11$\rm \arcsec/\rm pixel$. Observations were taken on UT2015-08-18 using a clear, blue-blocking (CBB) filter. These observations covered the second half of the eclipse, extending about 1 hour after egress. The observed eclipses from KELT, shown in Figure~\ref{fig:KELT} match well with the forward photodynamical models (Sec. \ref{sec:pd_model}).

To complement the {\it Kepler} data, we used the 1-m Mount Laguna Observatory telescope to observe \kic at the predicted times for planetary transits in the summer of 2015 (see Table~\ref{tab:CBP_transits}). Suboptimal observing conditions thwarted our efforts and, unfortunately, we were unable to detect the transits. The obtained images, however, confirmed the presence and orientation of the nearby star to the south of Kepler-1647.

\section{Unraveling the system}
\label{sec:pre_pd}

\begin{figure}
\centering
\epsscale{1.}
\plotone{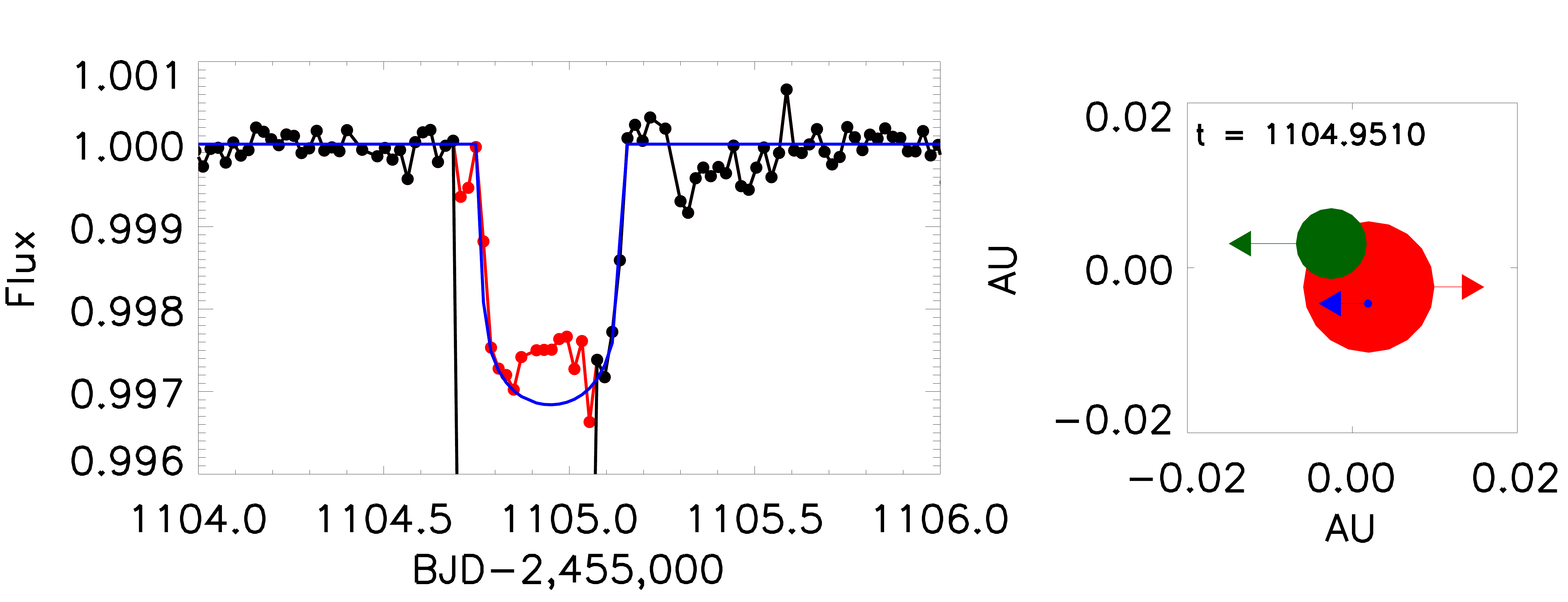}
\caption{The {\it Kepler} long-cadence light-curve (left panel) and the configuration (right panel) of the \kic system at the time of the syzygy shortly after a primary stellar eclipse. The planet does not cross the disk of the secondary star --  the configuration of the system and the sizes of the objects on the right panel are to scale, and the arrows indicate their sky-projected direction of motion. The red symbols on the left panel represent the light-curve after removal of a primary eclipse template; the blue curve on the left panel represents the fit to the transit of the CBP across the primary star. The suboptimal fit to the depth of the planetary transit is due to imperfect removal of the stellar eclipse caused by noisy data. 
\label{fig:blended_transit}}
\end{figure}

\kicb produced three transits: two across the secondary star at $t_{\rm B,1} = -3.0018$ and  $t_{\rm B, 2} = 1109.2612$ (BJD-2,455,000), corresponding to EB phase of 0.60 and 0.39 respectively, and one heavily blended transit across the primary star at time $t_{\rm A, 1}$ -- a syzygy (the secondary star and the planet simultaneously cross the line of sight between \kepler~and the disk of the primary star) during a primary stellar eclipse at $t_{\rm prim} = 1104.8797$. The latter transit is not immediately obvious and requires careful inspection of the light-curve. We measured transit durations across the secondary of $t_{\rm dur, B, 1} = 0.137$ days and $t_{\rm dur, B, 2} = 0.279$ days. The light-curve and the configuration of the system at the time of the syzygy are shown in Figure~\ref{fig:blended_transit}. 

To extract the transit center time and duration across the primary star, $t_{\rm A}$ and $t_{\rm dur, A}$, we subtracted a primary eclipse template from the data, then measured the mid-transit time and duration from the residual light-curve -- now containing only the planetary transit -- to be $t_{\rm A} = 1104.9510$ (BJD-2,455,000), corresponding to EB phase 0.006, and $t_{\rm dur, A} = 0.411$ days. From the transit depths, and accounting from dilution, we estimated $k_{\rm p, prim} = R_{\rm p}/R_{\rm A} = 0.06$ and $k_{\rm p, sec} = R_{\rm p}/R_{\rm B} = 0.11$. We outline the parameters of the planetary transits in Table~\ref{tab:CBP_transits}. As expected from a CBP, the two transits across the secondary star vary in both depth and duration, depending on the phase of the EB at the respective transit times. This unique observational signature rules out common false positives (such as a background EB) and confirms the nature of the planet.  In the following section we describe how we used the center times and durations of these three CBP transits to analytically describe the planet's orbit. 

\subsection{Analytic treatment}
\label{sec:pl_transits}

As discussed in Schneider \& Chevreton (1990) and Kostov et al.~(2013), the detection of CBP transits across one or both stars during the same inferior conjunction (e.g.,~\keplerp-34, -35 and now \kic) can strongly constrain the orbital configuration of the planet when the host system is an SB2 eclipsing binary. Such scenario, dubbed ``double-lined, double-dipped'' (Kostov et al.~2013), is optimal in terms of constraining the a priori unknown orbit of the CBP. 

\subsubsection{CBP transit times}

Using the RV measurements of the EB (Table~\ref{tab: tab_RV}) to obtain the {\it x-} and {\it y-}coordinates of the two stars at the times of the CBP transits, and combining these with the measured time interval between two consecutive CBP transits, we can estimate the orbital period and semi-major axis of the CBP --  independent of and complementary to the photometric-dynamical model presented in the next Section. Specifically, the CBP travels a known distance $\Delta x$ for a known time $\Delta t$ between two consecutive transits across either star (but during the same inferior conjunction), and its {\it x-}component velocity is\footnote{The observer is at $+z$, and the sky is in the $xy$-plane.}:

\begin{equation}
\begin{split}
V_{x,\rm p} = \frac{\Delta x}{\Delta t} \\
\Delta x = | x_i - x_j |\\
\Delta t = | t_i - t_j |
\end{split}
\label{eq:deltaV}
\end{equation}

{\noindent where $x_{i,j}$ is the {\it x}-coordinate of the star being transited by the planet at the observed mid-transit time of $t_{i,j}$}. As described above, the CBP transited across the primary and secondary star during the same inferior conjunction, namely at $t_i = t_{\rm A} = 1104.9510\pm0.0041$, and at $t_j = t_{\rm B,2} = 1109.2612\pm0.0036$ (BJD-2,455,000). Thus $\Delta t = 4.3102\pm0.0055$ days.  

The barycentric {\it x}-coordinates of the primary and secondary stars are (Hilditch, 2001)\footnote{The longitude of ascending node of the binary star, $\Omega_{\rm bin}$, is undefined and set to zero throughout this paper.}:

\begin{equation}
x_{\rm A,B}(t_{\rm A,B}) = r_{\rm A,B}(t_{\rm A,B})\cos[\theta_{\rm A,B}(t_{\rm A,B}) + \omega_{\rm bin}]
\label{eq:xAB}
\end{equation}
 
{\noindent where $r_{\rm A,B}(t_{\rm A,B})$ is the radius-vector of each star at the times of the CBP transits $t_{\rm A}$ and $t_{\rm B,2}$:}

\begin{equation}
r_{\rm A,B}(t_{\rm A,B}) = a_{\rm A,B}(1-e_{\rm bin}^{2})[1 + e_{\rm bin}\cos(\theta_{\rm A,B}(t_{\rm A,B}))]^{-1}
\label{eq:rAB}
\end{equation}

{\noindent where $\theta_{\rm A,B}(t_{\rm A,B})$ are the true anomalies of the primary and secondary stars at $t_{\rm A}$ and $t_{\rm B}$, $e_{\rm bin} = 0.159\pm0.003$ and $\omega_{\rm bin}=300.85\pm0.91\arcdeg$ are the binary eccentricity and argument of periastron as derived from the spectroscopic measurements (see Table~\ref{tab: tab_RV})}. Using the measured semi-amplitudes of the RV curves for the two host stars ($K_{\rm A} = 55.73\pm0.21$~km/sec and $K_{\rm B} = 69.13\pm0.5$~km/sec, see Table 1), and the binary period $P_{\rm bin} = 11.2588$ days, the semi major axes of the two stars are $a_{\rm A} = 0.0569\pm0.0002$~AU and $a_{\rm B} = 0.0707\pm0.0005$~AU respectively (Equation 2.51, Hilditch (2001)). 

To find $\theta_{\rm A,B}(t_{\rm A,B})$ we solved Kepler's equation for the two eccentric anomalies\footnote{Taking into account that $\omega$ for the primary star is $\omega_{\rm bin}-\pi$}, $E_{\rm A,B}-e_{\rm bin}\sin(E_{\rm A,B}) = 2\pi(t_{\rm A,B} - t_{\rm 0})/P_{\rm bin}$, where $t_{\rm 0} = -47.869\pm0.003~(\rm BJD - 2,455,000)$ is the time of periastron passage for the EB, and obtain $\theta_{\rm A}(t_{\rm A})=2.64\pm0.03$ rad, $\theta_{\rm B}(t_{\rm B,2})= 4.55\pm0.03$ rad. The radius vector of each star is then $r_{\rm A}(t_{\rm A}) = 0.065\pm0.003$~AU and $r_{\rm B}(t_{\rm B}) = 0.071\pm0.004$~AU, and from Equation \ref{eq:xAB}, $x_{\rm A}(t_{\rm A}) = 0.0020\pm0.0008$~AU and $x_{\rm B}(t_{\rm B}) = -0.0658\pm0.0009$~AU. Thus $\Delta x = 0.0681\pm0.0012$ AU and, finally, $V_{x,\rm p} = 0.0158\pm0.0006$~AU/day (see Equation \ref{eq:deltaV}).

Next, we used $V_{x,\rm p}$ to estimate the period and semi-major axes of the CBP as follows. The {\it x}-component of the planet's velocity, assuming that $\cos(\Omega_{\rm p}) = 1$ and $\cos(i_{\rm p}) = 0$ where $\Omega_{\rm p}$ and $i_{\rm p}$ are the planet's longitude of ascending node and inclination respectively in the reference frame of the sky)\footnote{Both consistent with the planet transiting near inferior conjunction.} is:

\begin{equation}
\begin{split}
V_{\rm x,p} =  - \left(\frac{2 \pi G M_{\rm bin}}{P_{\rm p}}\right)^{1/3} \frac{[e_{\rm p}\sin\omega_{\rm p} + \sin(\theta_{\rm p} + \omega_{\rm p})]}{\sqrt{(1-e_{\rm p}^{2})}}
\end{split}
\label{eq:vel}
\end{equation}

{\noindent where $M_{\rm bin} = 2.19\pm0.02~\Msun$ is the mass of the binary star (calculated from the measured radial velocities of its two component stars -- Equation 2.52, Hilditch 2001), and $P_{\rm p}, e_{\rm p}, \omega_{\rm p}$ and $\theta_{\rm p}$ are the orbital period, eccentricity, argument of periastron and the true anomaly of the planet. When the planet is near inferior conjunction, like during the transits at $t_{\rm A}$ and $t_{\rm B,2}$, we can approximate $\sin(\theta_{\rm p} + \omega_{\rm p})=1$. Simple algebra shows that:

\begin{equation}
\begin{aligned}
P_{\rm p}=1080~\frac{(1 + e_{\rm p}\sin\omega_{\rm p})^3}{(1-e_{\rm p}^{2})^{3/2}}~[\rm days]
\label{eq:cbp_period}
\end{aligned}
\end{equation}

{\noindent Thus if the orbit of \kicb is circular, its orbital period is $P_{\rm p}\approx1100$ days and its semi-major axis is $a_{\rm p}\approx2.8$~AU.} In this case we can firmly rule out a CBP period of 550 days.

Even if the planet has a non-zero eccentricity, Equation \ref{eq:cbp_period} still allowed us to constrain the orbit it needs to produce the two transits observed at $t_{\rm A}$ and $t_{\rm B,2}$. In other words, if $P_{\rm p}$ is indeed not $\sim1100$ days but half of that (assuming that a missed transit fell into a data gap), then from Equation \ref{eq:cbp_period}:

\begin{equation}
\begin{split}
2~(1 + e_p\sin\omega_p )^3 &= (1-e_p^{2})^{3/2} \leq 1\\
\end{split}
\label{eq:inequalities}
\end{equation}

{\noindent which implies ${e_{\rm p}} \geq 0.21$.} Thus Equation \ref{eq:inequalities} indicates that unless the eccentricity of CBP \kicb is greater than 0.21, its orbital period cannot be half of 1100 days. 

We note that our Equation \ref{eq:inequalities} differs from Equation 8 in Schneider \& Chevreton (1990) by a factor of $4\pi^3$. As the two equations describe the same phenomenon (for a circular orbit for the planet), we suspect there is a missing factor of $2\pi$ in the sine and cosine parts of their equations 6a and 6b, which propagated through.

\subsubsection{CBP transit durations}

As shown by Kostov et al.~(2013, 2014) for the cases of \keplerp-47b, \keplerp-64b and \keplerp-413b, and discussed by Schneider \& Chevreton (1990), the measured transit durations of CBPs can constrain the a priori unknown mass of their host binary stars\footnote{Provided they are single-lined spectroscopic binaries.} when the orbital period of the planets can be estimated from the data. The case of \kic is the opposite -- the mass of the EB is known (from spectroscopic observations) while the orbital period of the CBP cannot be pinned down prior to a full photodynamical solution of the system. However, we can still estimate the orbital period of \kicb using its transit durations:

\begin{equation}
t_{\rm dur,~\it n} = \frac{2~R_{\rm c,~\it n}}{V_{x, \rm p} + V_{x, \rm star, \it n}}
\label{eq:durations}
\end{equation}

{\noindent where $R_{\rm c,\it n} = R_{\rm star, \it n}\sqrt{(1+k_{\rm p})^2 - b_{n}^2}$ is the transit chord (where $k_{\rm p,prim} = 0.06$, $k_{\rm p,sec}=0.11$, and $b_{n}$ the impact parameter) for the $n-th$ CBP transit, $V_{x, \rm star, \it n}$ and $V_{x, \rm p}$ are the x-component velocities of the star and of the CBP respectively.} Using Equation \ref{eq:vel}, we can rewrite Equation \ref{eq:durations} in terms of $P_{\rm p}$ (the orbital period of the planet) near inferior conjunction ($\sin(\theta_{\rm p} + \omega_{\rm p}) = 1$) as:

\begin{equation}
P_{\rm p}(t_{\rm dur,\it n}) = 2 \pi G M_{\rm bin}(1 + e_{\rm p}\sin\omega_{\rm p})^{3}{\left(\frac{2R_{\rm c,\it n}}{t_{\rm dur,\it n}}-V_{x,\rm star,\it n}\right)^{-3}{(1-e_{\rm p}^{2})}^{-3/2}} \\
\label{eq:cbp_per_dur}
\end{equation}

{\noindent where the stellar velocities $V_{x, \rm star,\it n}$ can be calculated from the observables (see Equation 3, Kostov et al.~2013): $V_{x,\rm A} = 2.75\times10^{-2}$~AU/day; $V_{x,\rm B,1} = 4.2\times10^{-2}$~AU/day; and $V_{x,\rm B,2} = 2\times10^{-2}$~AU/day. Using the measured values for $t_{\rm dur, A},~t_{\rm dur, B, 1}$, $t_{\rm dur, B, 2}$ (listed in Table~\ref{tab:CBP_transits}), and requiring $b_n\geq0$, for a circular orbit of the CBP we obtained:

\begin{equation}
\begin{split}
P_{\rm p}(t_{\rm dur,~A})~[\rm days]\geq 4076\times\left(2.4~(\frac{R_{\rm A}}{\Rsun}) - 2.75\right)^{-3} \\
P_{\rm p}(t_{\rm dur,~B,~1})~[\rm days]\geq 4076\times\left(4.1~(\frac{R_{\rm A}}{\Rsun}) - 4.2\right)^{-3} \\
P_{\rm p}(t_{\rm dur,~B,~2})~[\rm days]\geq4076\times\left(2~(\frac{R_{\rm A}}{\Rsun}) - 2\right)^{-3}
\end{split}
\label{eq:cbp_dur}
\end{equation}

We show these inequalities, defining the allowed region for the period of the CBP as a function of the (a priori uncertain) primary radius, in Figure~\ref{fig:durations}. The allowed regions for the CBP period are above each of the solid lines (green for Transit A, Red for Transit B2, and blue for Transit B1), where the uppermost line provides the strongest lower limit for the CBP period at the specific $R_{\rm A}$. The dotted vertical lines denote, from left to right, the radius provided by NexSci ($R_{\rm A, NexSci} = 1.46~\Rsun$), calculated from our photodynamical model ($R_{\rm A, PD} = 1.79~\Rsun$, Sec \ref{sec:pd_model}), and inferred from the rotation period ($R_{\rm A, rot} = 2.08~\Rsun$). As seen from the figure, $R_{\rm A, NexSci}$ is too small as the corresponding minimum CBP periods are too long -- if this was indeed the primary stellar radius then the measured duration of transit A (the strongest constraint at that radius) would correspond to $P_{\rm CBP}\sim10^4$ days (for a circular orbit). The rotationally-inferred radius is consistent with a planet on either a 550-day orbit or a 1100-day orbit but given its large uncertainty (see Section \ref{sec:rot}) the constraint it provides on the planet period is rather poor.  

Thus while the measured transit durations cannot strictly break the degeneracy between a 550-day and an 1100-day CBP period prior to a full numerical treatment, they provide useful constraints. Specifically, without any prior knowledge of the transit impact parameters, and only assuming a circular orbit, Equation \ref{eq:cbp_per_dur} indicates that a) the impact parameter of transit B1 must be large (in order to bring the blue curve close to the other two); and b) $R_{\rm A}$ must be greater than $\sim1.75~\Rsun$ (where the green and red lines intersect) so that the three inequalities in Equation \ref{eq:cbp_dur} are consistent with the data. 

We note that $P_{\rm p}$ in Equation \ref{eq:cbp_per_dur} is highly dependent on the impact parameter of the particular CBP transit (which is unknown prior to a full photodynamical solution). Thus a large value for $b_{\rm B,1}$ will bring the transit B1 curve (blue) closer to the other two. Indeed, our photodynamical model indicates that $b_{\rm B,1}\approx0.7$, corresponding to $P_{\rm p}(t_{\rm dur,~B,~1})=1105$ days from Equation \ref{eq:cbp_per_dur} (also blue arrow in Figure~\ref{fig:durations}) -- bringing the blue curve in line with the other two and validating our analytic treatment of the transit durations. The other two impact parameters, $b_{\rm A,1}$ and $b_{\rm B,2}$, are both $\approx0.2$ according to the photodynamical model, and do not significantly affect Equation~\ref{eq:cbp_per_dur} since their contribution is small (i.e., $b_{\rm A}^2\approx b_{\rm B,2}^2\approx0.04$). For the photodynamically-calculated $R_{\rm A, PD} = 1.79~\Rsun$, the respective analytic CBP periods are $P_{\rm p}(t_{\rm dur,~A})=1121$~days and $P_{\rm p}(t_{\rm dur,~B,~2})=1093$~days. Thus the analytic analysis presented here is fully consistent with the comprehensive numerical solution of the system -- which we present in the next section. 

\begin{figure}
\centering
\plotone{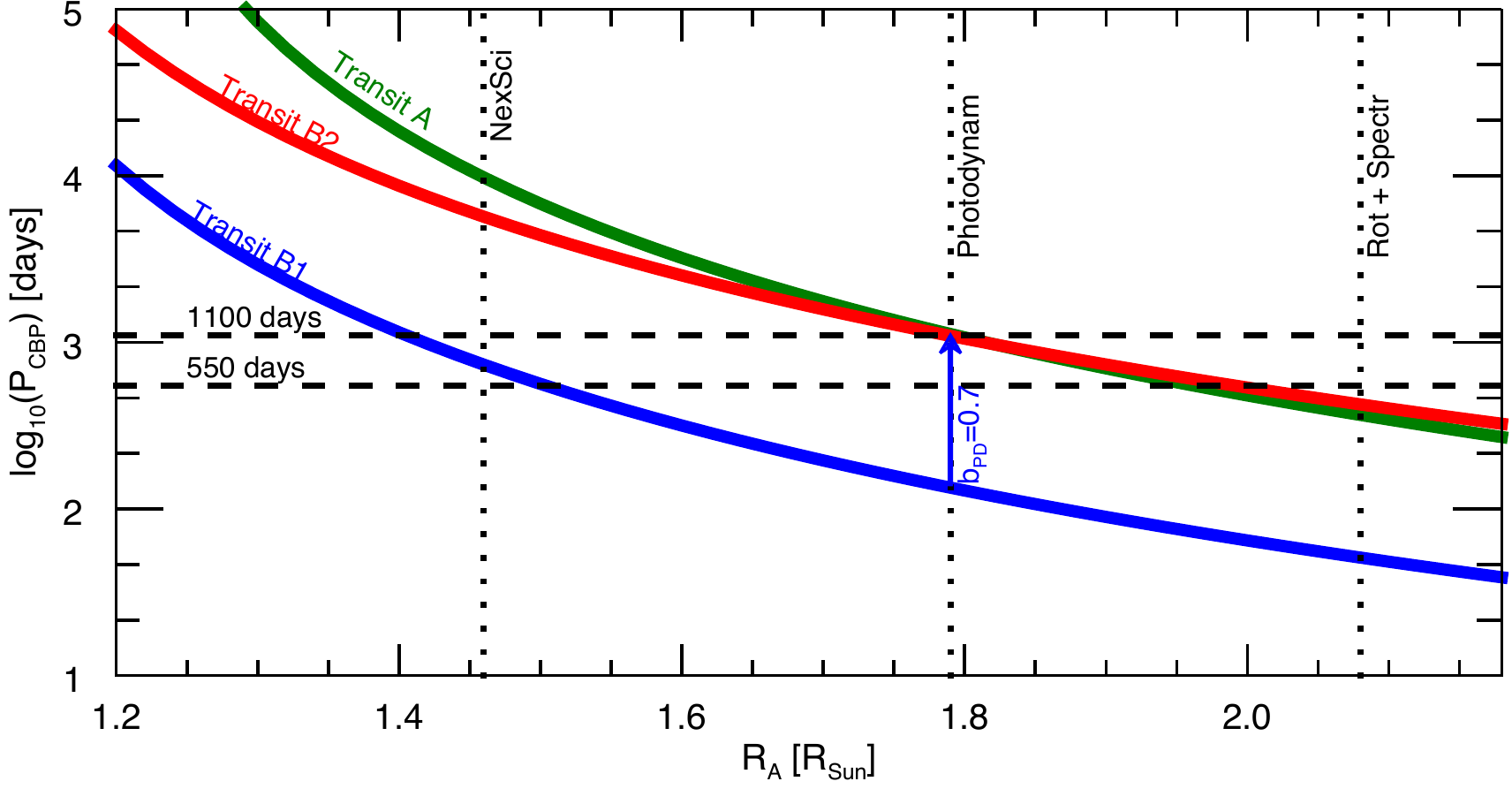}
\caption{Analytic constraints on the minimum allowed period of the CBP (the region above each solid line, Equation \ref{eq:cbp_dur}), as a function of the measured transit durations and the a priori uncertain primary radius $(R_{\rm A})$, and assuming a circular CBP orbit. The colors of the solid lines correspond to Transit A (green), Transit B2 (red), and Transit B1 (blue). The uppermost curve for each radius constrains the minimum period the most. From left to right, the dotted vertical lines represent the primary radius provided by NexSci, calculated from our photodynamical model, and inferred from our rotation and spectroscopy analysis. To be consistent with the data,  $R_{\rm A}$ must be larger than $\sim1.75~\Rsun$, and the impact parameter of transit B1 must be large -- in line with the photodynamical solution of the system. Accounting for the photodynamically-measured impact parameter for transit B1 (blue arrow) makes the analytically-derived period of the planet fully consistent with the numerically-derived value of $\sim 1100$ days.
\label{fig:durations}}
\end{figure}

\subsection{Photometric-dynamical solution}
\label{sec:pd_model}

CBPs reside in dynamically-rich, multi-parameter space where a strictly Keplerian solution is not adequate. A comprehensive description of these systems requires a full photometric-dynamical treatment based on the available and follow-up data, on N-body simulations, and on the appropriate light-curve model. We describe this treatment below.

\subsubsection{Eclipsing light-curve (ELC)}

To obtain a complete solution of the \kic system we used the ELC code (Orosz \& Hauschildt 2000) with recent ``photodynamical'' modifications (e.g.,~Welsh et al.~2015). The code fully accounts for the gravitational interactions between all bodies. Following Mardling \& Lin (2002), the gravitational force equations have been modified to account for precession due to General Relativistic (GR) effects and due to tides.  

Given initial conditions (e.g.,~masses, positions, and velocities for each body), the code utilizes a 12$^{th}$ order symplectic Gaussian Runge-Kutta integrator (Hairer \& Hairer 2003) to calculate the 3-dimensional positions and velocities of the two stars and the planet as a function of time. These are combined with the light-curve model of Mandel \& Agol (2002), and the quadratic law limb-darkening coefficients, using the ``triangular'' sampling of Kipping (2013), to calculate model light and radial velocity curves which are directly compared to the {\it Kepler} data (both long- and, where available, short-cadence), the measured stellar radial velocities, and the ground-based
light-curves.

The ELC code uses the following as adjustable parameters. The three masses and three sizes of the occulting objects, the Keplerian orbital elements (in terms of Jacobian coordinates) for the EB and the CBP ($e_{\rm bin}, e_{\rm p}, i_{\rm bin}, i_{\rm p}, \omega_{\rm bin}, \omega_{\rm p}, P_{\rm bin}, P_{\rm p}$,
and times of conjunction $T_{\rm c, bin}$, $T_{\rm c,p}$)\footnote{The time of conjunction is defined as the conjunction with the barycenter of the system. For the EB, this is close to a primary eclipse while the CBP does not necessarily transit at conjunction.}, the CBP longitude of ascending node $\Omega_{\rm p}$ ($\Omega_{\rm bin}$ is undefined, and set to zero throughout), the ratio between the stellar temperatures, the quadratic limb-darkening coefficients of each star in the {\em Kepler} bandpass and primary limb darkening coefficients for the three bandpasses used for the ground-based observations, and the seasonal contamination levels of the {\it Kepler} data. The GR modifications to the force equations require no additional parameters, and the modifications to account for apsidal motion require the so-called $k_2$ constant and the ratio of the rotational frequency of the star to its pseudosynchronous value for each star. For fitting purposes, we used parameter composites or ratios (e.g.,~$e\cos\omega,~e\sin\omega, ~M_{\rm A}/M_{\rm B},~R_{\rm A}/R_{\rm B}$) as these are generally better constrained by the data. We note that the parameters quoted here are the instantaneous``osculating'' values. The coordinate system is Jacobian, so the orbit of the planet is referred to the center-of-mass of the binary star. These values are valid for the reference epoch only since the orbits of the EB and the CBP evolve with time, and must be used in the context of a dynamical model to reproduce our results.

As noted earlier, star spot activity is evident in the \kepler light-curve.  After some initial fits, it was found that some of the eclipse profiles were contaminated by star spot activity. We carefully examined the residuals of the fit and selected five primary and five secondary eclipses that have ``clean'' residuals (these are shown in Figure~\ref{fig:eclipses}).  We fit these clean profiles, the times of eclipse for the remaining eclipse events (corrected for star spot contamination), the three ground-based observations, and the two radial velocity curves.  We used the observed rotational velocities of each star and the spectroscopically determined flux ratio between the two stars (see Tables \ref{tab: tab_RV} and \ref{tab:the_EB}) as additional constraints. The model was optimized using a Differential Evolution Monte Carlo Markov Chain (DE-MCMC, ter Braak 2006).  A total of 161 chains were used, and 31,600 generations were computed. We skipped the first 10,000 generations for the purposes of computing the posterior distributions.  We adopted the parameters from the best-fitting model, and use posterior distributions to get the parameter uncertainties. 

Throughout the text we quote the best-fit parameters. These do not have error bars and, as noted above, should be interpreted strictly as the parameters reproducing the light-curve. For uncertainties the reader should refer to the mode and the mean values calculated from the posterior distributions. 
 
\subsubsection{Consistency check}

We confirmed the ELC solution with the {\it photodynam} code (Carter et al.~2011, previously used for a number of CBPs, e.g.,~Doyle et al.~2011, Welsh et al.~2012, Schwamb et al.~2013, Kostov et al.~2014). We note, however, that the {\it photodynam} code does not include tidal apsidal motion -- which must be taken into account for \kic as discussed above. Thus while the solution of the {\it photodynam} code provides adequate representation of the light-curve of \kic (the differences are indistinguishable by eye), it is not the best-fit model in terms of chi-square statistics. We outline the input for the {\it photodynam} code\footnote{In terms of osculating parameters.} required to reproduce the light-curve of \kic in Table~\ref{tab:pd_in}.

We also carried out an analysis using an independent photodynamical code (developed by co-author D.R.S.) that is based on the Nested Sampling concept. This code includes both GR and tidal distortion. Nested Sampling was introduced by Skilling (2004) to compute the Bayesian Evidence (marginal likelihood or the normalizing factor in Bayes' Theorem). As a byproduct, a representative sample of the posterior distribution is also obtained. This sample may then be used to estimate the statistical properties of parameters and of derived quantities of the posterior. MultiNest was an implementation by Feroz et al.~(2008) of Nested Sampling incorporating several improvements. Our version of MultiNest is based on the Feroz code and is a parallel implementation. The MultiNest solution confirmed the ELC solution. 

\section{Results and Discussion}
\label{sec:results}

The study of extrasolar planets is first and foremost the study of their parent stars, and transiting CBPs, in particular, are a prime example. As we discuss in this section\footnote{And also shown by the previous \kepler CBPs.}, their peculiar observational signatures, combined with {\it Kepler's} unmatched precision, help us to not only decipher their host systems by a comprehensive photodynamical analysis, but also constrain the fundamental properties of their host stars to great precision. In essence, transiting CBPs allow us to extend the ``royal road'' of EBs (Russell 1948) to the realm of exoplanets.

\subsection{The \kic system}

Our best-fit ELC photodynamical solution for \kic and the orbital configuration of the system are shown in Figure~\ref{fig:elc_lc} and \ref{fig:2939_orbit}, respectively. The correlations between the major parameters are shown in Figure~\ref{fig:fig_corr}. The ELC model parameters are listed in Tables \ref{tab:ELC_in} (fitting parameters), \ref{tab:ELC_out} (derived Keplerian elements) and \ref{tab:Barycentric_cartesian} (derived Cartesian). Table~\ref{tab:ELC_in} lists the mode and mean of each parameter as well as their respective uncertainties as derived from the MCMC posterior distributions; the upper and lower bounds represent the 68$\%$ range. The sub-percent precision on the stellar masses and sizes (see the respective values in Table~\ref{tab:ELC_out}) demonstrate the power of photodynamical analysis of transiting CB systems. The parameters presented in Tables \ref{tab:ELC_out} and \ref{tab:Barycentric_cartesian} represent the osculating, best-fit model to the \kepler light-curve, and are only valid for the reference epoch (BJD - 2,455,000). These are the parameters that should be used when reproducing the data. The mid-transit times, depths and durations of the observed and modeled planetary transits are listed in Table~\ref{tab:CBP_transits}.

The central binary \kic is host to two stars with masses of  $M_{\rm A} = \vMstarA\pm\eMstarA~\Msun$, and $M_{\rm B} = \vMstarB\pm\eMstarB~\Msun$, and radii of $R_{\rm A} =\vRstarA\pm\eRstarA~\Rsun$, and $R_{\rm B} = \vRstarB\pm\eRstarB~\Rsun$, respectively (Table~\ref{tab:ELC_out}). The temperature ratio of the two stars is $T_{\rm B}/T_{\rm A} = 0.9365\pm0.0033$ and their flux ratio is $F_{\rm B} /F_{\rm A} = 0.21\pm0.02$ in the {\it Kepler} band-pass. The two stars of the binary revolve around each other every 11.25882 days in an orbit with a semi-major axis of $a_{\rm bin} = \vabin\pm\eabin$~AU, eccentricity of $e_{\rm bin} = 0.1602\pm0.0004$, and inclination of $i_{\rm bin} = 87.9164\pm0.0145^\circ$ (see also Table~\ref{tab:ELC_out} for the respective mean and mode values).

\begin{figure}
\centering
\plotone{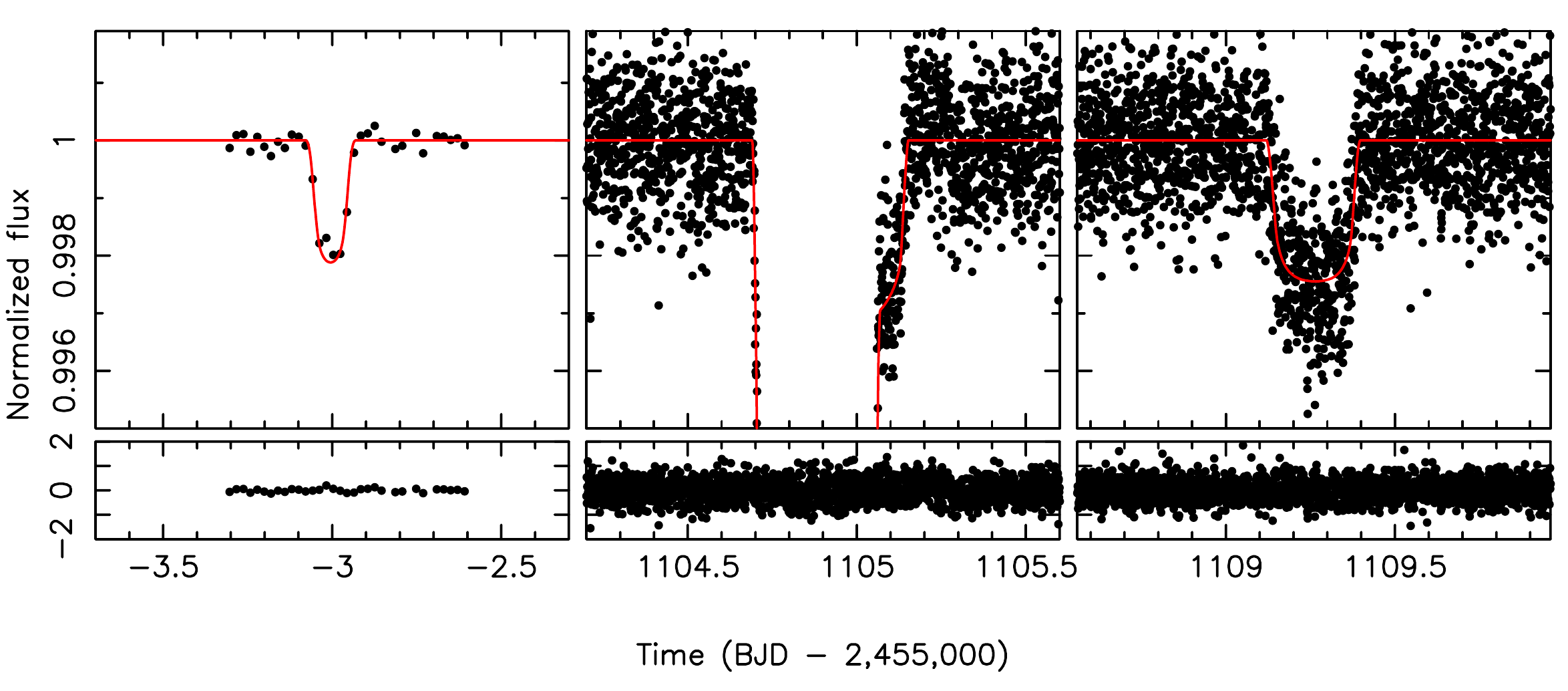}
\caption{ELC photodynamical solution (red, or grey color) for the normalized \kepler light-curve (black symbols) of \kiccomma, centered on the three transits of the CBP, and the respective residuals.  The left panel shows long-cadence data, the middle and right panels show short-cadence data. The first and third transits are across the secondary star, the second transit (heavily blended with a primary stellar eclipse off the scale of the panel) is across the primary star -- during a syzygy. The model represents the data well.
\label{fig:elc_lc}}
\end{figure}

\begin{figure}
\plotone{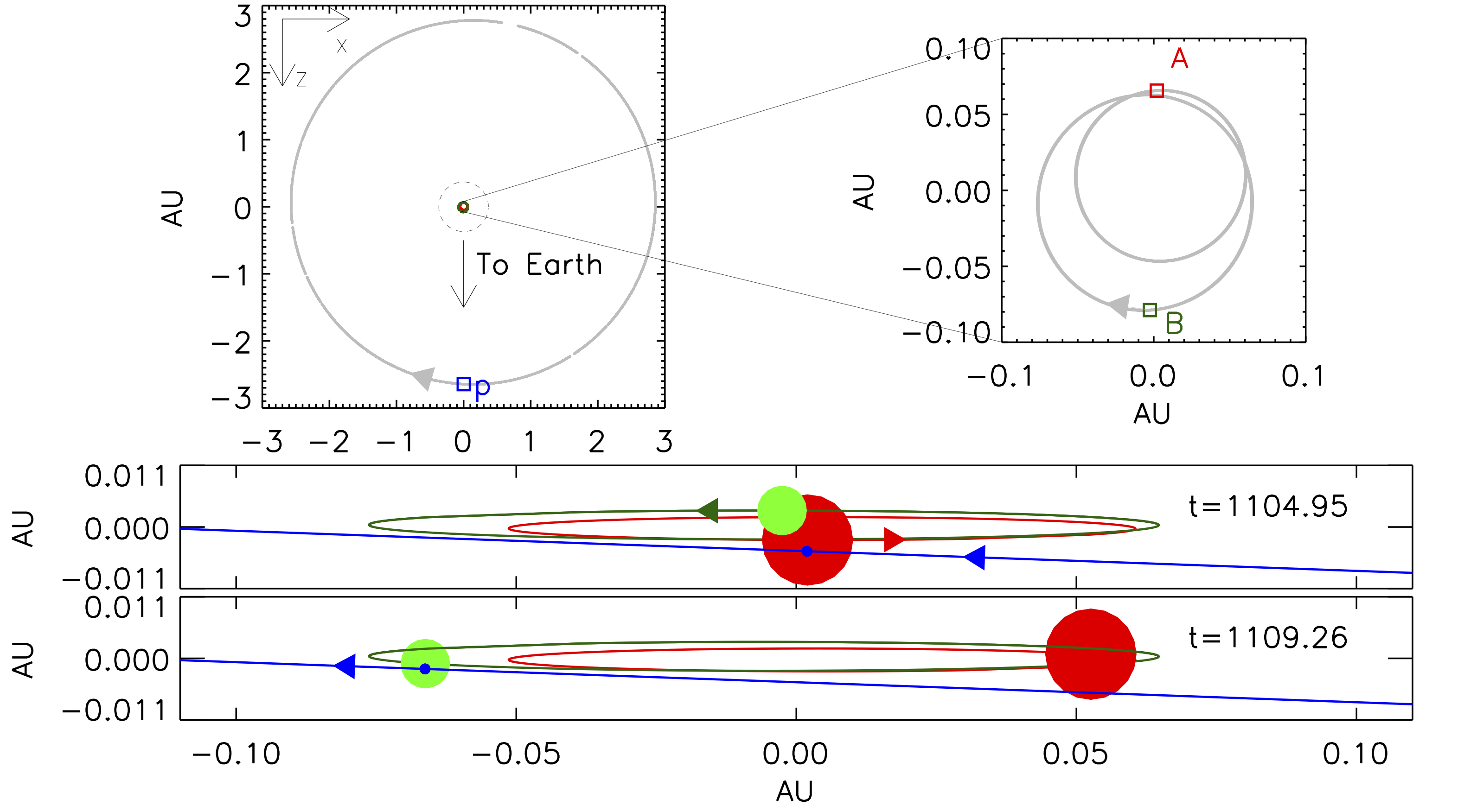}
\caption{ELC photodynamical solution for the orbital configuration of the \kicb system. The orbits and the symbols for the two stars in the lower two panels (red, or dark color for primary ``A'', green, or light color for the secondary ``B'') are to scale; the planet symbols in the lower two panels (blue color) are exaggerated by a factor of two for viewing purposes. The upper two panels show the configuration of the system at $t_{\rm A} = 1104.95$ (BJD - 2,455,000) as seen from above; the dashed line in the upper left panel represents the minimum distance from the EB for dynamical stability. The lower two panels show the configuration of the system (and the respective directions of motion) at two consecutive CBP transits during the same conjunction for the planet: at $t_{\rm A} = 1104.95$ and at $t_{\rm B,2} = 1109.26$ (BJD - 2,455,000). The orientation of the $xz$ coordinate axes (using the nomenclature of Murray \& Correia 2011) is indicated in the upper left corner of the upper left panel.
\label{fig:2939_orbit}}
\end{figure}

\begin{figure}
\centering
\plotone{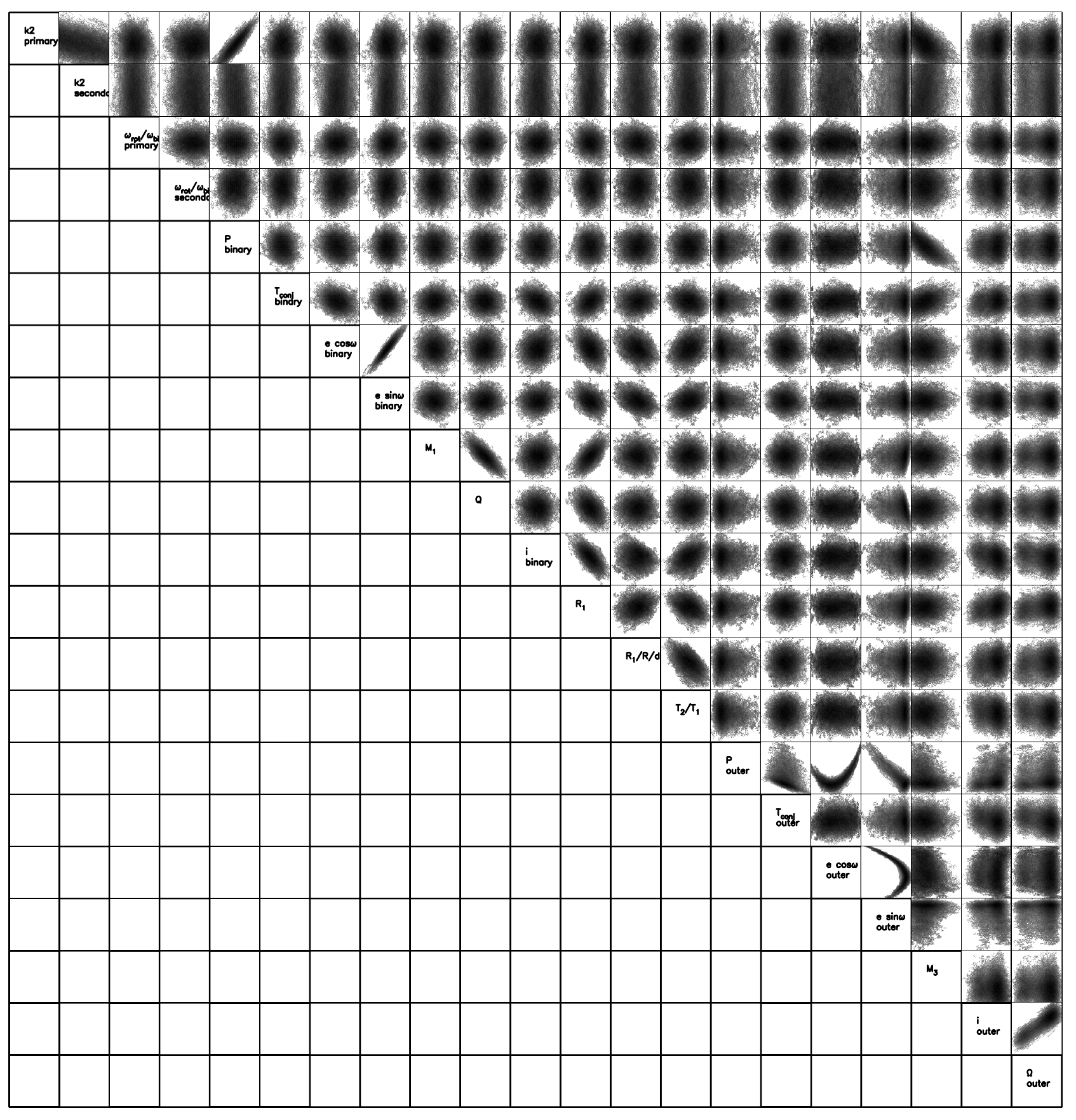}
\caption{Correlations between the major parameters for the ELC photodynamical solution. 
\label{fig:fig_corr}}
\end{figure}

The accurate masses and radii of the \kic stars, along with our constraints on the temperatures and metallicity of this system enable a useful comparison with stellar evolution models. As described in Section~3.2, a degeneracy remains in our determination of [m/H] and $T_{\rm {\rm eff}}$, which we resolved here by noting that 1) current models are typically found to match the observed properties of main-sequence F stars fairly well \citep[see, e.g.,~][]{Torres:10}, and 2) making the working assumption that the same should be true for the primary of \kiccomma, of spectral type approximately F8 star. We computed model isochrones from the Yonsei-Yale series \citep{Yi01, Demarque04} for a range of metallicities. In order to properly compare results from theory and observation, at each value of metallicity, we also adjusted the spectroscopic temperatures of both stars by interpolation in the table of $T_{\rm eff,A}$ and $T_{\rm eff,B}$ versus [m/H] mentioned in Section~3.2. This process led to an excellent fit to the primary star properties (mass, radius, temperature) for a metallicity of ${\rm [m/H]} = -0.14\pm0.05$ and an age of $4.4\pm0.25$~Gyr. This fit is illustrated in the mass-radius and mass-temperature diagrams of Figure~\ref{fig:RvM}. We found that the temperature of the secondary is also well fit by the same model isochrone that matches the primary. The radius of the secondary, however, is only marginally matched by the same model, and appears nominally larger than predicted at the measured mass. Evolutionary tracks for the measured masses and the same best-fit metallicity are shown in Figure~\ref{fig:logg_v_T}.

One possible cause for the slight tension between the observations and the models for the secondary in the mass-radius diagram is a bias in either the measured masses or the radii. While the individual masses may indeed be subject to systematic uncertainty, the mass \emph{ratio} should be more accurate, and a horizontal shift in the upper panels of Figure~\ref{fig:RvM} can only improve the agreement with the secondary at the expense of the primary. Similarly, the sum of the radii is tightly constrained by the photodynamical fit, and the good agreement between the spectroscopic and photometric estimates of the flux ratio (see Table~\ref{tab:the_EB}) is an indication that the radius ratio is also accurate. The flux ratio is a very sensitive indicator because it is proportional to the square of the radius ratio.

An alternative explanation for this tension may be found in the physical properties of the secondary star. As discussed in Section~\ref{sec:rot}, \kic \,B is an active star with a rotation period of 11 days. In addition, the residuals from the times of eclipse (Figure~\ref{fig:spot_slope}) provide evidence of spots on the surface of this star, likely associated with strong magnetic fields. Such stellar activity is widely believed to be responsible for ``radius inflation'' among stars with convective envelopes \citep[see, e.g.][]{Torres:13}. The discrepancy between the measured and predicted radius for the radius of \kic \,B amounts to 0.014~$\Rsun$ (1.4\%), roughly 2.5 times the radius uncertainty at a fixed mass.  Activity-induced radius inflation also generally causes the temperatures of late-type stars to be too cool compared to model predictions. This ``temperature suppression'' is, however, not seen in the secondary star of \kic possibly because of systematic errors in our spectroscopic temperatures or because the effect may be smaller than our uncertainties. We note that while radius and temperature discrepancies are more commonly seen in M dwarfs, the secondary of \kic is not unique in showing some of the same effects despite being much more massive (only $\sim$3\% less massive than the Sun). Other examples of active stars of similar mass as \kic \,B with evidence of radius inflation and sometimes temperature suppression include V1061\,Cyg\,B \citep{Torres:06}, CV\,Boo\,B \citep{Torres:08}, V636\,Cen\,B \citep{Clausen:09}, and EF\,Aqr\,B \citep{Vos:12}. In some of these systems, the effect is substantially larger than that in \kic \,B, and can reach up to 10\%.

We note that while the best-fit apsidal constant for the primary star is consistent with the corresponding models of Claret et al.~(2006) within one sigma\footnote{For 4.4 Gyr, and Z = 0.01 -- closest to the derived age and metallicity of Kepler-1647.} (i.e.~$k_{2}(\rm A, ELC) = 0.003\pm0.005$ vs $k_{2}(\rm A, Claret, M=1.2~\Msun) = 0.004$), the nominal uncertainties on the secondary's constant indicate a tension (i.e.~$k_{2}(\rm B, ELC) = 0.03\pm0.0005$ vs $k_{2}(\rm B, Claret, M=1.0~\Msun) = 0.015$). To evaluate the significance of this tension on our results, we repeat the ELC fitting as described in the previous section, but fixing the apsidal constants to those from Claret et al.~(2006), i.e.~$k_{2}(\rm A)=0.004$ and $k_{2}(\rm B)=0.015$. The fixed-constants solution is well within one sigma of the solution presented in Tables \ref{tab:ELC_in} and \ref{tab:ELC_out} (where the constants are fit for), and the planet's mass remains virtually unchanged, i.e.~${\rm M_p(best-fit, fixed~k_2) = 462\pm167~M_{Earth}}$ vs ${\rm M_p(best-fit, free~k_2) = 483\pm206~M_{Earth}}$.

\begin{figure}
\centering
\epsscale{0.9}
\plotone{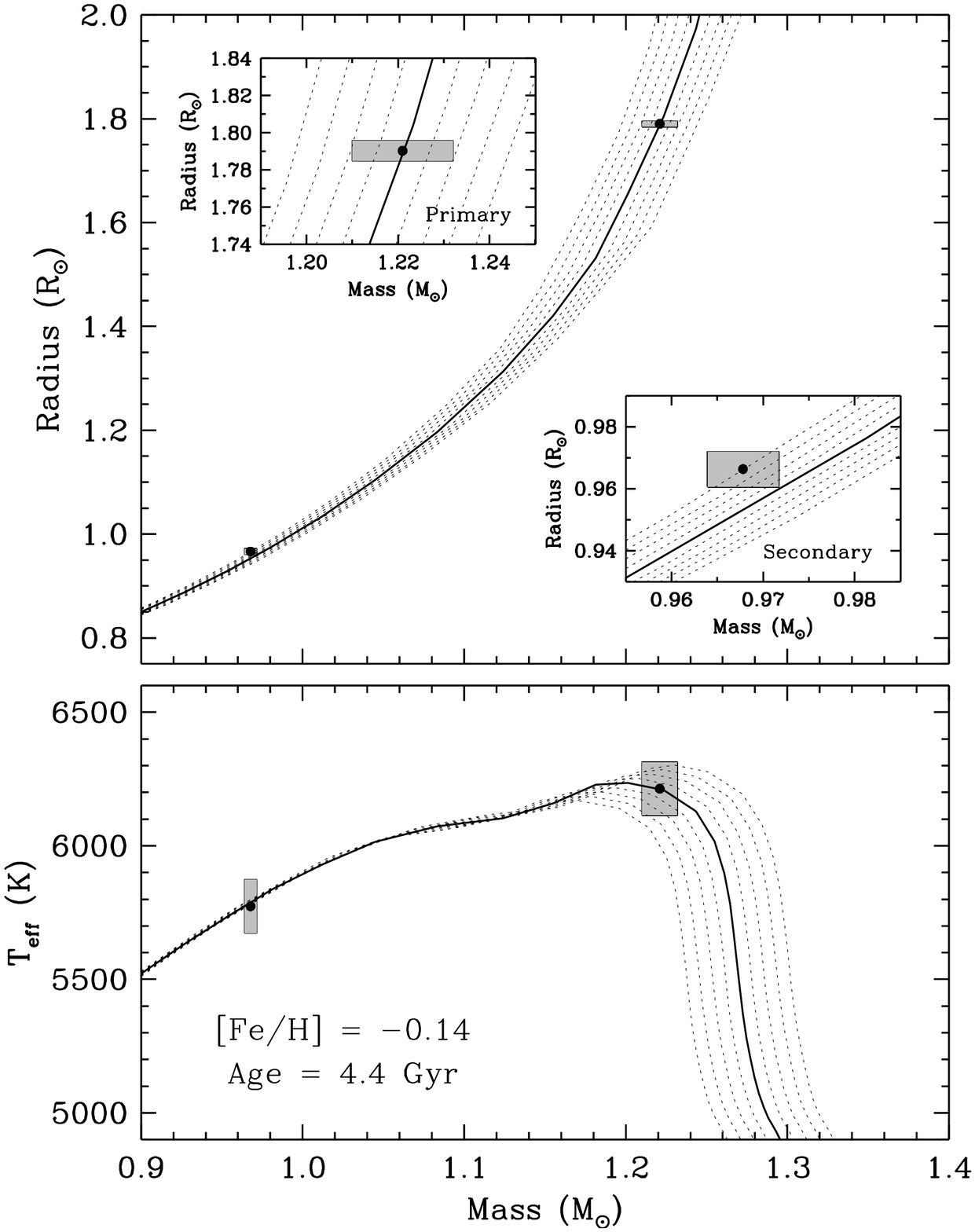}
\caption{Radii (top panel) and temperatures (bottom) of the \kic components as a function of mass, compared with stellar evolution models from the Yonsei-Yale series \citep{Yi01, Demarque04}. Model isochrones are shown for the best-fit metallicity of ${\rm [m/H]} = -0.14$ and ages of 4.0--4.8~Gyr in steps of 0.1~Gyr, with the solid line marking the best-fit age of 4.4~Gyr. The insets in the top panel show enlargements around the primary and secondary observations.
\label{fig:RvM}}
\end{figure}

\begin{figure}
\centering
\plotone{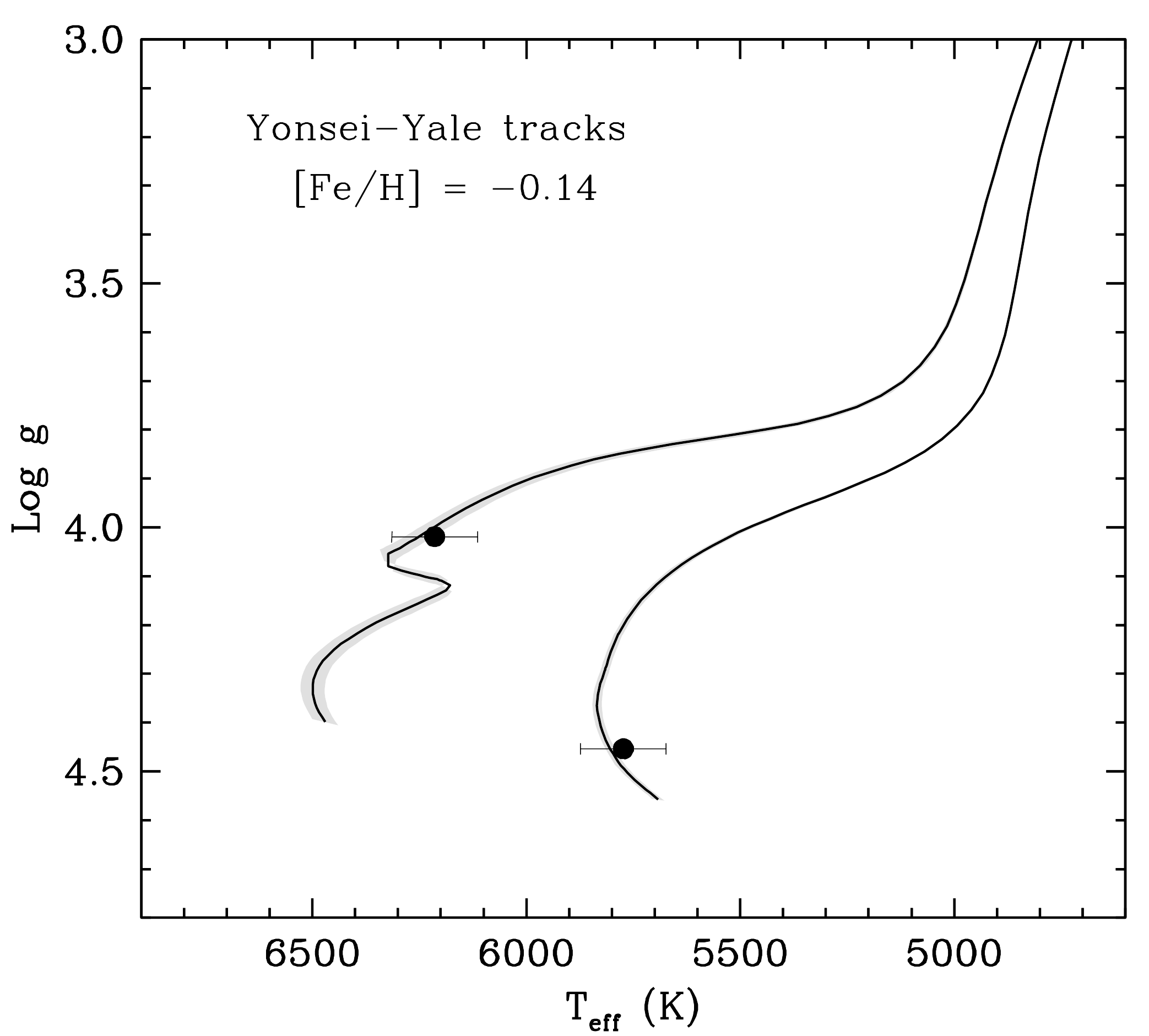}
\caption{Measurements for \kic in the $\log g$--$T_{\rm eff}$ plane, along with 4.4~Gyr evolutionary tracks from the Yonsei-Yale series calculated for the measured masses and a metallicity of ${\rm [m/H]} = -0.14$. The shaded area around each track indicates the uncertainty in its location stemming from the mass errors. The left track is for the primary star, the right track for the secondary star. 
\label{fig:logg_v_T}}
\end{figure}

\subsection{The Circumbinary Planet}

Prior to the discovery of \kicbcomma, all the known \kepler CBPs were Saturn-sized and smaller (the largest being \keplerp-34b, with a radius of $0.76~\Rjup$), and were found to orbit their host binaries within a factor of two from their corresponding boundary for dynamical stability. Interestingly, {\it ``Jupiter-mass CBPs are likely to be less common because of their less stable evolution''}, suggest PN08, and {\it ``if present [Jupiter-mass CBPs] are likely to orbit at large distances from the central binary.''}. Indeed -- at the time of writing, with a radius of $\vRplanetjup\pm\eRplanetjup$~\Rjup~($\vRplanetear\pm\eRplanetear~R_{\rm Earth}$) and mass of $\vMplanetjup\pm\eMplanetjup$~\Mjup~($\vMplanetear\pm\eMplanetear$~$M_{\rm Earth}$)\footnote{See also Table~\ref{tab:ELC_out} for mean and mode.}, \kicb is the first Jovian transiting CBP from \kepler, and the one with the longest orbital period ($P_p = 1107.6$ days). The size and the mass of the planet are consistent with theoretical predictions indicating that substellar-mass objects evolve towards the radius of Jupiter after $\sim1$~Gyr of evolution for a wide range of initial masses ($\sim0.5-10~\Mjup$), and regardless of the initial conditions (`hot' or `cold' start) (Burrows et al.~2001, Spiegel \& Burrows 2013). To date, \kicb is also one of the longest-period transiting planets, demonstrating yet again the discovery potential of continuous, long-duration observations such as those made by {\it Kepler}. The orbit of the planet is nearly edge-on ($i_{\rm p} = 90.0972\pm0.0035^\circ$), with a semi-major axis $a_{\rm p} = \vaplanet\pm\eaplanet$~AU and eccentricity of $e_{\rm p} = 0.0581\pm0.0689$.

\subsection{Orbital Stability and Long-term Dynamics}
\label{sec:stability}

Using Equation 3 from Holman \& Weigert (1999) (also see Dvorak~1986, Dvorak et al.~1989), the boundary for orbital stability around \kic is at $a_{crit} = 2.91~a_{\rm bin}$. With a semi-major axis of 2.72~AU ($21.3~a_{\rm bin}$), \kicb is well beyond this stability limit indicating that the orbit of the planet is long-term stable. To confirm this, we also integrated the planet-binary system using the best-fit ELC parameters for a timescale of 100 million years. The results are shown in Figure~\ref{fig:long_term_100M}. As seen from the figure, the semi-major axis and eccentricity of the planet do not experience appreciable variations (that would inhibit the overall orbital stability of its orbit) over the course of the integration. 

On a shorter timescale, our numerical integration indicate that the orbital planes of the binary and the planet undergo a 7040.87-years, anti-correlated precession in the plane of the sky. This is illustrated in Figure~\ref{fig:com_precession}. As a result of this precession, the orbital inclinations of the binary and the planet oscillate by $\sim0.0372^{\circ}$ and $\sim2.9626^{\circ}$, respectively (see upper and middle panels in Figure~\ref{fig:com_precession}). The mutual inclination between the planet and the binary star oscillates by $\sim0.0895^{\circ}$, and the planet's ascending node varies by $\sim2.9665^{\circ}$ (see middle and lower left panels in Figure~\ref{fig:long_term_100M}). 

Taking into account the radii of the binary stars and planet, we found that CBP transits are possible only if the planet's inclination varies between $89.8137^{\circ}$ and $90.1863^{\circ}$. This transit window is represented by the red horizontal lines in the middle panel of Figure~\ref{fig:com_precession}. The planet can cross the disks of the two stars only when the inclination of its orbit lies between these two lines. To demonstrate this, in the lower panel of Figure~\ref{fig:com_precession} we expand the vertical scale near 90 degrees, and also show the impact parameter $b$ for the planet with respect to the primary (solid green symbols) and secondary (open blue symbols). Transits only formally occur when the impact parameter, relative to the sum of the stellar and planetary radii, is less than unity. We computed the impact parameter at the times of transit as well as at every conjunction, where a conjunction is deemed to have occurred if the projected separation of the planet and star is less than 5 planet radii along the $x$-coordinate\footnote{The $xy$-coordinates define the plane of the sky, and the observer is along the $z$-coordinate.}. Monitoring the impact parameter at conjunction allowed us to better determine the time span for which transits are possible as shown in Figure~\ref{fig:com_precession}. These transit windows, bound between the horizontal red lines in Figure~\ref{fig:com_precession}, span about 204 years each. As a result, over one precession cycle, planetary transits can occur for $\approx408$ years (spanning two transit windows, half a precession cycle apart), i.e., $\approx5.8\%$ of the time. The transits of the CBP will cease in $\sim160$~years.

Following the methods described in Welsh et al.~(2012), we used the probability of transit for \kicb to roughly estimate the frequency of \kic-like systems. At the present epoch, the transit probability is approximately 1/300 (i.e.,~$R_{\rm A}/a_{\rm p}\sim0.33\%$), where the enhancement due to the barycentric motion of an aligned primary star is a minimal ($\sim5\%$) effect. Folding in the probability of detecting two transits at two consecutive conjunctions of an 1100-day period planet when observing for 1400-days (i.e.,~ 300/1400, or $\sim20\%$), the probability of detecting \kicb is therefore about $0.33\%\times20\%$, or approximately 1/1500. Given that 1 such system was found out of Kepler's $\sim150,000$ targets, this suggests an occurrence rate of roughly 1\%. A similar argument can be used to analyze the frequency among Eclipsing Binaries in particular. As the population of EBs is already nearly aligned to the line-of-sight, the probability of alignment for the CBP is significantly increased. In particular, as described in Welsh et al. (2012), the probability that the planet will be aligned given that the target is an EB is approximately ${\rm a_{bin}/a_p} = 0.046$~(compared to $\sim0.003$ for the isotropic case). Recent results suggest that only EBs with periods longer than $\sim7$ days contain CBPs (Welsh et al. 2013; Martin et al.~2015; Mu\~{n}oz \& Lai~2015, Hamers et al.~2015); there are $\sim1000$ such Kepler EBs (Kirk et al.~2015). Thus, if 1\% of these 1000 EBs have Kepler-1647-like CBPs, then the expected number of detections would be $0.02\times1000\times0.046 = 0.92\approx1$. As a result, the Kepler-1647 system suggests that $\sim2\%$ of all eligible EBs have similar planets.

Cumming et al.~(2008) suggest a $\sim1\%$ occurrence rate of Jupiter mass planets within the 2-3 AU range. Therefore, the relative frequency of such planets around FGK main sequence stars is similar for both single and binary stars. This interesting result is consistent with what has been found for other CBPs as well and certainly has implications for planet formation theories. A more detailed analysis following the methods of Armstrong et al.~(2014) and Martin \& Triaud~(2015) in context of the full CBP population is beyond the scope of the present work. 

\begin{figure}
\centering
\plotone{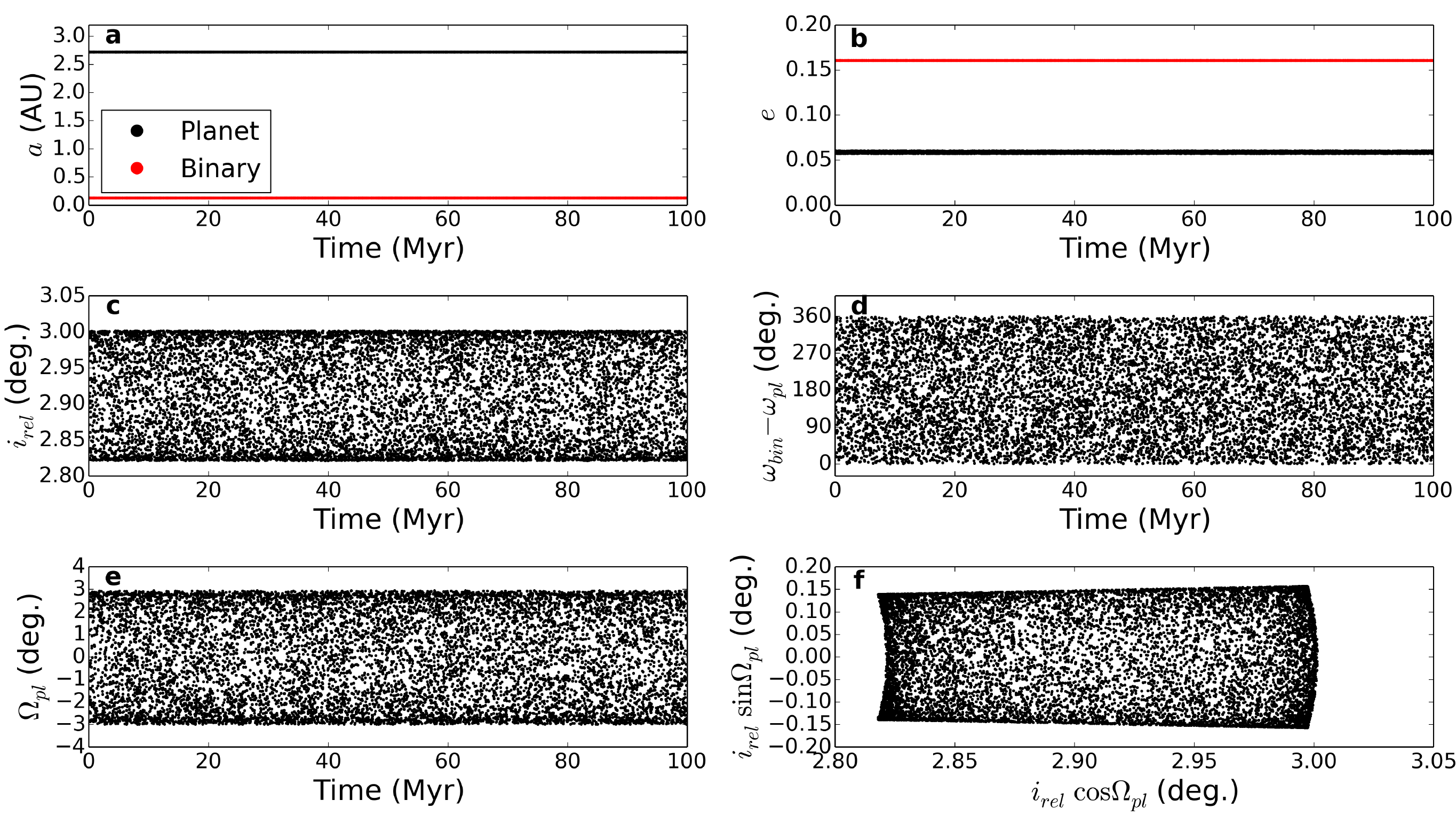}
\caption{The long-term (100 Myr) evolution of select orbital elements of the \kic system. We do not observe chaotic behavior confirming that the CBP is long-term stable.
\label{fig:long_term_100M}}
\end{figure}

\begin{figure}
\centering
\epsscale{1.}
\plotone{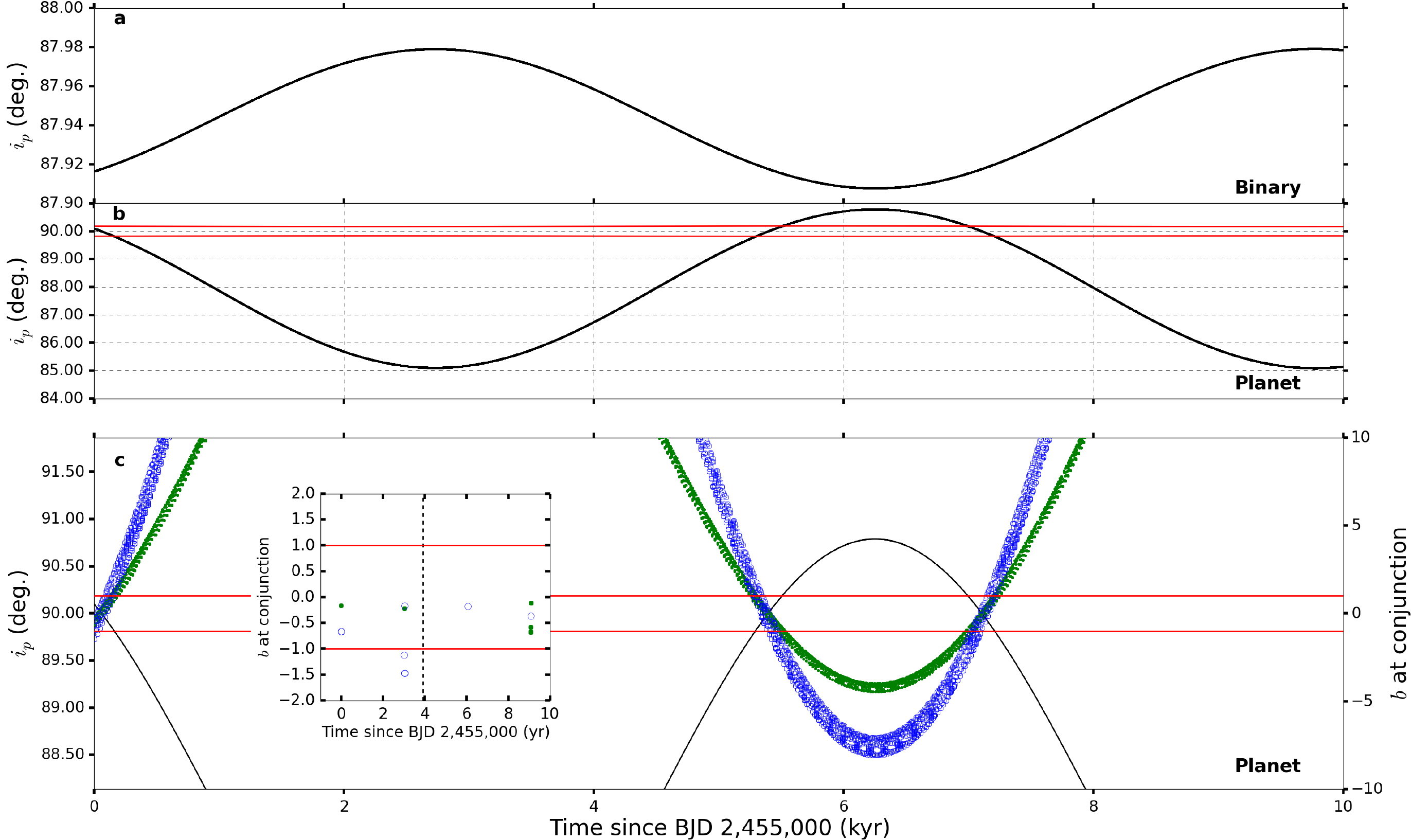}
\caption{Precession of the orbital inclination (in the reference frame of the sky) of the binary star (upper panel) and of the CBP (middle panel) over 10,000 years, and evolution of the impact parameter ($b$) between the planet and each star (lower panel). The solid, red horizontal lines in the middle panel represent the windows where planetary transits are possible. Each transit window is $\approx204$ years, indicating that the CBP can transit for $\approx5.8\%$ of its precession cycle. The lower panel is similar to the middle, but zoomed around the red horizontal band and showing the two impact parameters near primary (solid green symbols) and secondary (open blue symbols) conjunction. The inset in the lower panel is zoomed around the duration of the \kepler mission, with the vertical dashed line indicating the last data point of Quarter 17. There were four transits over the duration of the mission, with one of them falling into a data gap (green symbol near time 0 in the inset panel). 
\label{fig:com_precession}}
\end{figure}

We performed a thorough visual inspection of the light-curve for additional transits. Our search did not reveal any obvious features.\footnote{A single feature reminiscent of a very shallow transit-like event can be seen near day 1062 (BJD - 2,455,000) but we could not associate it with the planet.}. However, given than \kicb is far from the limit for orbital stability, we also explored whether a hypothetical second planet, interior to \kicbcomma, could have a stable orbit in the \kic system, using the Mean Exponential Growth factor of Nearby Orbits (MEGNO) formalism (Cincotta et al.~2003, Go{\'z}dziewski et al. 2001, Hinse et al.~2015). Figure~\ref{fig:MEGNO} shows the results of our simulations in terms of a 2-planet MEGNO map of the \kic system. The map has a resolution of Nx=500 \& Ny=400, and was produced from a set of 20,000 initial conditions. The x and y axes of the figure represent the semi-major axis and eccentricity of the hypothetical second planet (with the same mass as \kicb). Each initial condition is integrated for 2,738 years, corresponding to 88,820 binary periods. A given run is terminated when the MEGNO factor $Y$ becomes larger than 12. The map in Figure~\ref{fig:MEGNO} shows the MEGNO factor in the interval $1 <Y< 5$. Quasi-periodic orbits at the end of the integration have ($|\langle Y\rangle - 2.0| < 0.001$). The orbital elements used for creating this map are Jacobian geometric elements, where the semi-major axis of the planet is relative to the binary center of mass. All interactions (including planet-planet perturbation) are accounted for, and the orbits of the planets are integrated relative to the binary orbital plane. The initial mean anomaly of the second hypothetical planet is set to be 180$^\circ$ away from that of \kicb (i.e.,~the second planet starts at opposition). Quasi-periodic initial conditions are color-coded purple in Figure~\ref{fig:MEGNO}, and chaotic dynamics is color-coded in yellow. As seen from the figure, there are regions where such a second planet may be able to maintain a stable orbit for the duration of the integrations. We further explored our MEGNO results by integrating a test orbit of the hypothetical second planet for $10^7$ years\footnote{Using an accurate adaptive-time step algorithm, http://www.unige.ch/$\sim$hairer/prog/nonstiff/odex.f}, and found it to be stable (for the duration of the integrations) with no significant onset of chaos. However, we note that the orbital stability of such hypothetical planet will change dramatically depending on its mass.

\begin{figure}
\centering
\plotone{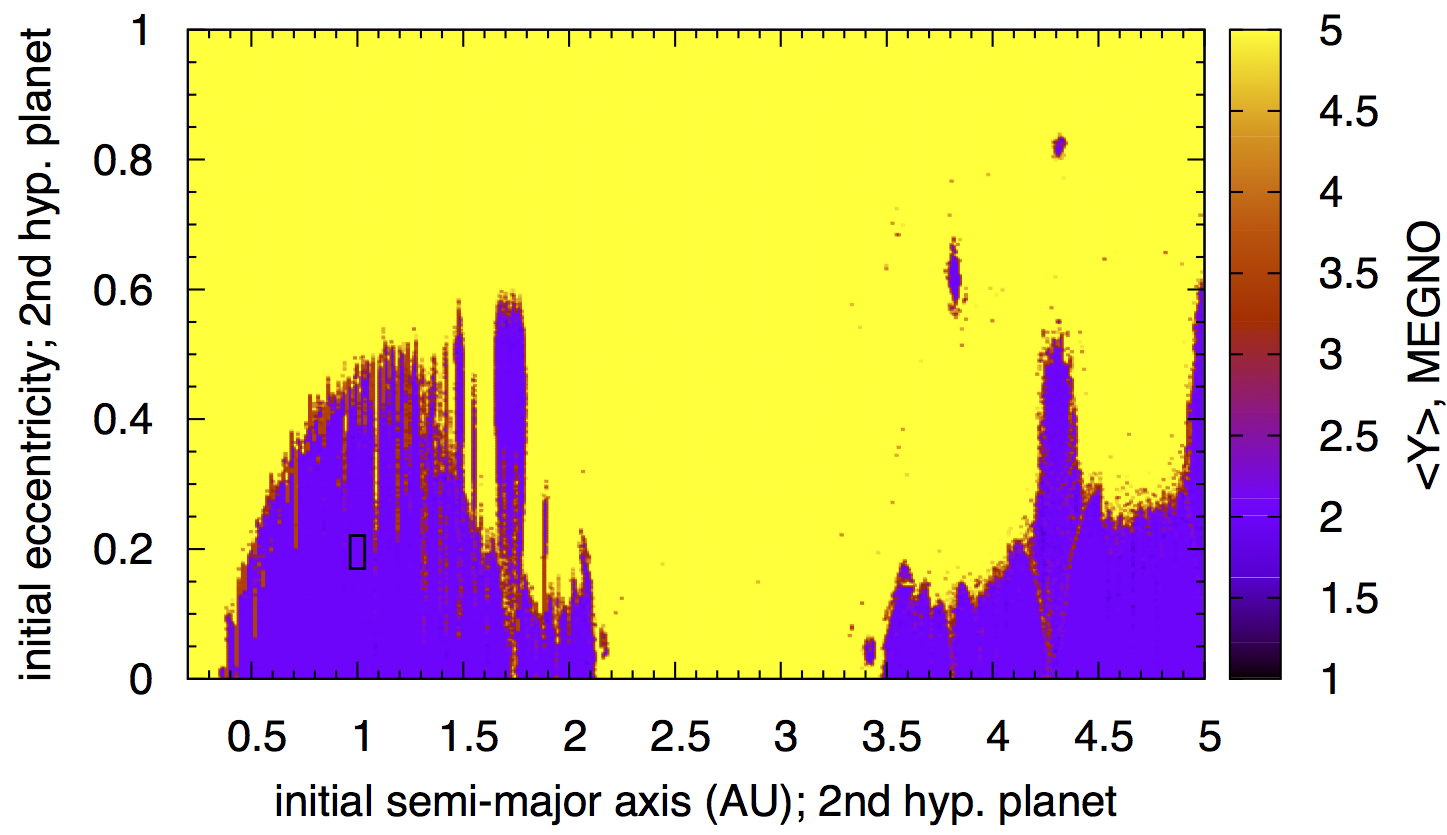}
\caption{Dynamical stability (in terms of the MEGNO $Y$ factor) for a hypothetical second planet, interior and coplanar to \kicbcomma, and with the same mass, as a function of its orbital separation and eccentricity.  The purple regions represent stable orbits for up to 2,738 years, indicating that there are plausible orbital configurations for such a planet. The MEGNO simulations reproduce 
well the critical dynamical stability limit at 0.37 AU. The square symbol represents the initial condition for a test orbit of a hypothetical second planet that we integrated for 10 million years, which we found to be stable.
\label{fig:MEGNO}}
\end{figure}

\subsection{Stellar Insolation and Circumbinary Habitable Zone}
\label{sec:HZ}

Using the formalism presented by Haghighipour~\&~Kaltenegger~(2013), we calculated the circumbinary Habitable Zone (HZ) around \kiccomma. Figure~\ref{fig:hz_orbit} shows the top view of the HZ and the orbital configuration of the system at the time of the orbital elements in Table~\ref{tab:ELC_out}. The boundary for orbital stability is shown in red and the orbit of the planet is in white. The light and dark green regions represent the extended and conservative HZs (Kopparapu et al.~2013, 2014), respectively. The arrow shows the direction from the observer to the system. As shown here, although it takes three years for \kicb to complete one orbit around its host binary star, our best-fit model places this planet squarely in the conservative HZ for the entire duration of its orbit. This makes \kicb the fourth of ten currently known transiting CBPs that are in the HZ of their host binary stars. 

Figure~\ref{fig:insolation} shows the combined and individual stellar fluxes reaching the top of the planet's atmosphere. The sharp drops in the left and middle panels of the figure are due to the stellar eclipses as seen from the planet; these are not visible in the right panel due to the panel's sampling (500 days). The time-averaged insolation experienced by the CBP is $0.71\pm0.06$ $S_{\rm Sun-Earth}$. We caution that showing the summation of the two fluxes, as shown in this figure, are merely for illustrative purposes. This summation cannot be used to calculate the boundaries of the HZ of the binary star system. Because the planet's atmosphere is the medium through which insolation is converted to surface temperature, the chemical composition of this atmosphere plays an important role. As a result, the response of the planet's atmosphere to the stellar flux received at the location of the planet strongly depends on the wavelengths of incident photons which themselves vary with the spectral type of the two stars. That means, in order to properly account for the interaction of an incoming photon with the atmosphere of the planet, its contribution has to be weighted based on the spectral type (i.e.,~the surface temperature) of its emitting star. It is the sum of the spectrally-weighted fluxes of the two stars of the binary that has to be used to determine the total insolation, and therefore, the boundaries of the system's HZ (Haghighipour~\&~Kaltenegger 2013). The green dashed and dotted lines in Figure~\ref{fig:insolation} show the upper and lower values of this weighted insolation. As shown in Figure~\ref{fig:insolation}, the insolation received at the location of the CBP's orbit is such that the planet is completely in its host binary's HZ. 

For completeness, we note that while the gas giant \kicb itself is not habitable, it can potentially harbor terrestrial moons suitable for life as we know it (e.g.,~Forgan~\&~Kipping 2013, Hinkel~\&~Kane 2013, Heller et al.~2014). There is, however, no evidence for such moons in the available data. Transit timing offsets due to even Gallilean-type moons would be less than ten seconds for this planet.

\begin{figure}
\centering
\plotone{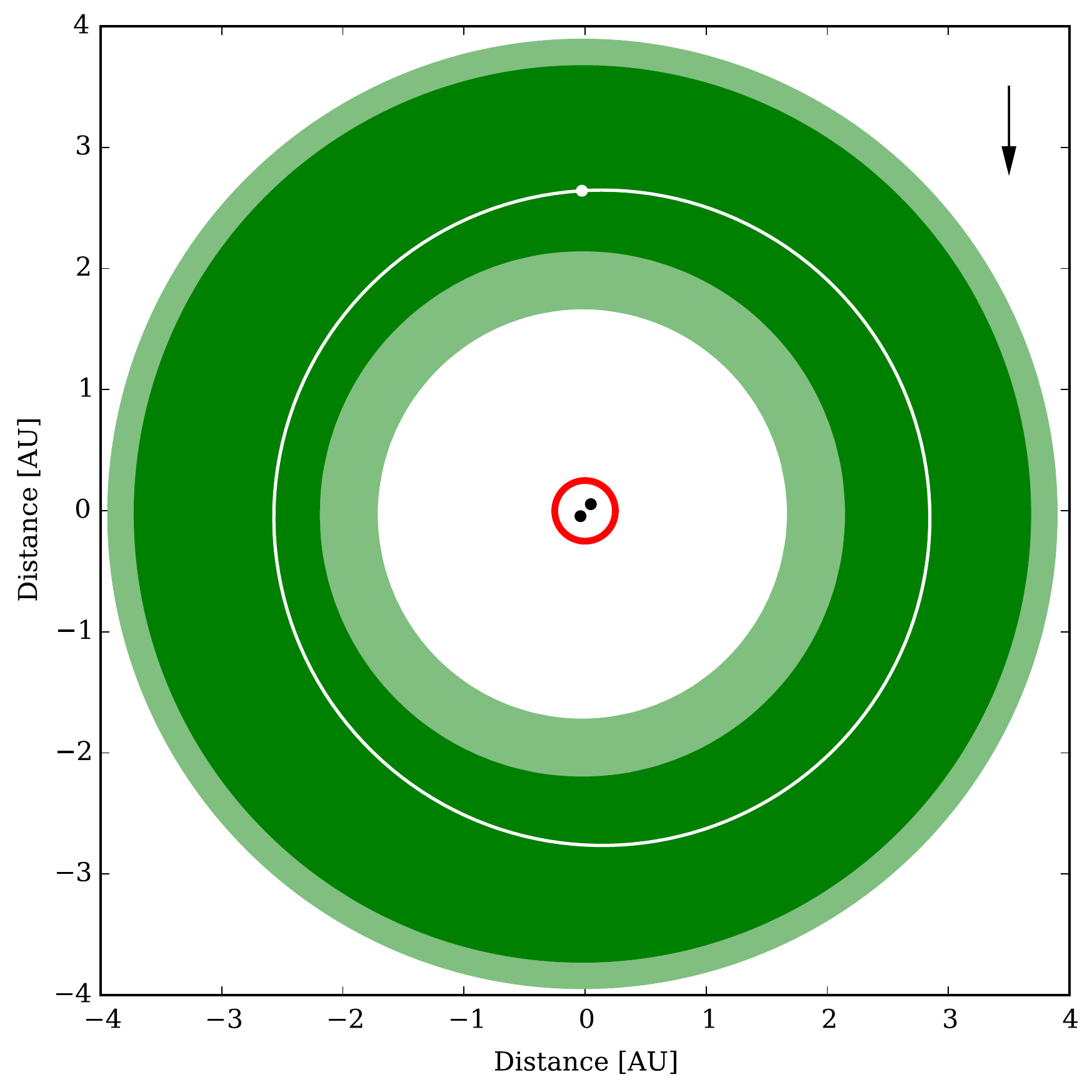}
\caption{Top view of the orbital configuration of the \kic system at time BJD = 2,455,000. The figure shows the location of the habitable zone (green) and the planet's orbit (white circle). The binary star is in the center, surrounded by the critical limit for dynamical stability (red circle). The dark and light green regions represent the conservative and extended HZ respectively (Kopparapu et al.~2013,~2014). The CBP is inside the conservative HZ for its entire orbit. The observer is looking from above the figure, along the direction of the arrow. A movie of the time evolution of the HZ can be found at http://astro.twam.info/hz-ptype .
\label{fig:hz_orbit}}
\end{figure}

\begin{figure}
\centering
\plotone{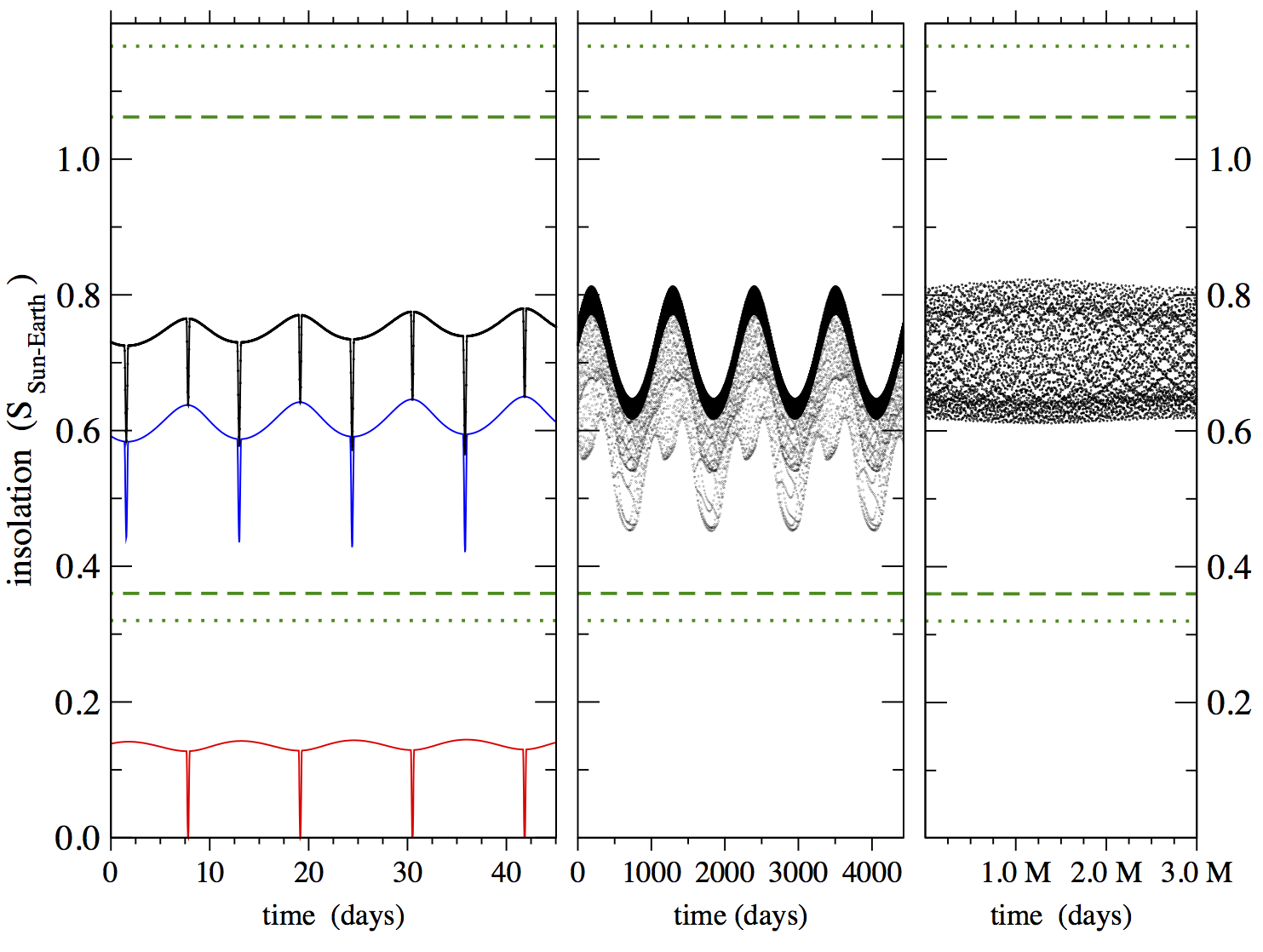}
\caption{Time evolution of the stellar flux (black curve) reaching the top of the CBP's atmosphere. The blue and red curves in the left panel represents the flux from the primary and secondary star respectively. The three panels represent the flux received by the CBP over four binary periods (left), four planetary periods (middle), and 3,000,000 days (right panel, with a sampling of every 500 days). The sharp features in the left panel represent the stellar eclipses (as seen from the planet); these are not present in the right panel due to the time sampling of the panel. The green dashed/dotted lines represent, from top to bottom, the runaway greenhouse, recent Venus, maximum greenhouse, and early Mars limits from Kopparapu et al.~(2013,~2014). The CBP is inside the conservative HZ for its entire orbit.
\label{fig:insolation}}
\end{figure}

\section{Conclusions}
\label{sec:conclusions}

We report the discovery and characterization of \kicb -- a new transiting CBP from the {\it Kepler} space telescope. In contrast to the previous transiting CBPs, where the planets orbit their host binaries within a factor of two of their respective dynamical stability limit, \kicb is comfortably separated from this limit by a factor of 7. The planet has an orbital period of $\sim$1100 days and a radius of $\vRplanetjup\pm\eRplanetjup~\Rjup$.  At the time of writing, \kicb is the longest-period, largest transiting CBP, and is also one of the longest-period transiting planets. With $\vMplanetjup\pm\eMplanetjup~\Mjup$, this CBP is massive enough to measurably perturb the times of the stellar eclipses. \kicb completed a single revolution during {\it Kepler's} observations and transited three times, one of them as a syzygy. We note that the next group of four transits (starting around date BJD = 2,458,314.5, or UT of 2018, July, 15, see Table~\ref{tab:CBP_transits}) will fall within the operation timeframe of {\it TESS}. The orbit of this CBP is long-term stable, with a precession period of $\sim7,040$ years. Due to its orbital configuration, \kicb can produce transits for only $\approx5.8\%$ of its precession cycle. Despite having an orbital period of three years, this planet is squarely in the conservative HZ of its binary star for the entire length of its orbit. 

The stellar system consists of two stars with $M_{\rm A} = \vMstarA\pm\eMstarA~\Msun$, $R_{\rm A} =\vRstarA\pm\eRstarA~\Rsun$, and $M_{\rm B} = \vMstarB\pm\eMstarB~\Msun$, $R_{\rm B} = \vRstarB\pm\eRstarB~\Rsun$ on a nearly edge-on orbit with an eccentricity of $\vebin\pm\eebin$. The two stars have a flux ratio of $F_{\rm B}/F_{\rm A} = 0.21\pm0.02$,  the secondary is an active star with a rotation period of $P_{\rm rot} = 11.23\pm0.01$ days, and the binary is in a spin-synchronized state. The two stars have effective temperatures of $T_{\rm A, \it eff} = 6210\pm100$~K and $T_{\rm B, \it eff} = 5770\pm125$~K respectively, metallicity of [Fe/H] = $-0.14\pm0.05$, and age of $4.4\pm0.25$~Gyr.

As important as a new discovery of a CBP is to indulge our basic human curiosity about distant worlds, its main significance is to expand our understanding of the inner workings of planetary systems in the dynamically rich environments of close binary stars. The orbital parameters of CBPs, for example, provide important new insight into the properties of protoplanetary disks and shed light on planetary formation and migration in the dynamically-challenging environments of binary stars. In particular, the observed orbit of \kicb lends strong support to the models suggesting that CBPs form at large distances from their host binaries and subsequently migrate either as a result of planet-disk interaction, or planet-planet scattering (e.g.,~Pierens \& Nelson~2013, Kley \& Haghighipour~2014, 2015).

\acknowledgments

We thank the referee for the insightful comments that helped us improve this paper. We thank Gibor Basri and Andrew Collier Cameron for helpful discussions regarding stellar activity, Michael Abdul-Masih, Kyle Conroy and Andrej Pr{\v s}a for discussing the photometric centroid shifts. This research used observations from the \kepler~Mission, which is funded by NASA's Science Mission Directorate; the TRES instrument on the Fred L. Whipple Observatory 1.5-m telescope; the Tull Coude Spectrograph on the McDonald Observatory 2.7-m Harlan J. Smith Telescope; the HIRES instrument on the W. M. Keck Observatory 10-m telescope; the HamSpec instrument on the Lick Observatory 3.5-m Shane telescope; the WHIRC instrument on the WIYN 4-m telescope; the Swarthmore College Observatory 0.6-m telescope; the Canela's Robotic Observatory 0.3-m telescope. This research made use of the SIMBAD database, operated at CDS, Strasbourg, France; data products from the Two Micron All Sky Survey (2MASS), the United Kingdom Infrared Telescope (UKIRT); the NASA exoplanet archive NexSci\footnote{http://exoplanetarchive.ipac.caltech.edu} and the NASA Community Follow-Up Observation Program (CFOP) website, operated by the NASA Exoplanet Science Institute and the California Institute of Technology, under contract with NASA under the Exoplanet Exploration Program. VBK and BQ gratefully acknowledge support by an appointment to the NASA Postdoctoral Program at the Goddard Space Flight Center and at the Ames Research Center, administered by Oak Ridge Associated Universities through a contract with NASA. WFW, JAO, GW, and BQ gratefully acknowledge support from NASA via grants NNX13AI76G and NNX14AB91G. NH acknowledges support from the NASA ADAP program under grant NNX13AF20G, and NASA PAST program grant NNX14AJ38G. TCH acknowledges support from KASI research grant 2015-1-850-04. Part of the numerical computations have been carried out using the SFI/HEA Irish Center for High-End Computing (ICHEC) and the POLARIS computing cluster at the Korea Astronomy and Space Science Institute (KASI). This work was performed in part under contract with the Jet Propulsion Laboratory (JPL) funded by NASA through the Sagan Fellowship Program executed by the NASA Exoplanet Science Institute. 

\bibliographystyle{apj}

\newpage

\clearpage
\begin{table}[ht]
\footnotesize
\begin{center}
\caption{Measured radial velocities.
\label{tab: tab_RV}}
\begin{tabular}{l|ll|ll}
\hline
\hline
BJD$_{\rm UTC}$ & $RV_{\rm A}$  & $\pm$$1\,\sigma_{\rm A}$ & $RV_{\rm B}$ & $\pm$$1\,\sigma_{\rm B}$ \\
$-2,455,000$ & (km\,s$^{-1}$) & (km\,s$^{-1}$) & (km\,s$^{-1}$) & (km\,s$^{-1}$) \\
\hline
 843.686310 (T)$^\dagger$ & 72.36 & 0.35 & -49.40 & 1.13 \\
 845.737695 (T) & 26.66 & 0.34 & 4.36 & 1.11 \\
 847.691513 (T) & -14.69 & 0.35 & 57.63 & 1.14 \\
 849.698453 (T) & -32.58 & 0.34 & 82.50 & 1.11 \\
 851.685281 (T) & 4.56 & 0.35 & 35.05 & 1.14 \\
 855.705432 (T) & 59.01 & 0.62 & -30.46 & 2.02 \\
 856.631747 (T) & 35.98 & 0.36 & -5.33 & 1.18 \\
 856.698255 (T) & 34.41 & 0.43 & -3.66 & 1.39 \\
1051.968336 (T) & -32.47 & 0.34 & 81.02 & 1.11 \\
1076.890883 (T) & 5.54 & 0.33 & 34.40 & 1.07 \\
1079.943631 (T) & 75.13 & 0.34 & -53.85 & 1.09 \\
1083.935311 (T) & -11.28 & 0.34 & 54.03 & 1.10 \\
1086.775726 (T) & -28.78 & 0.34 & 76.87 & 1.11 \\
1401.968283 (T) & -30.52 & 0.49 & 75.03 & 1.60 \\
 838.762187 (M)$^\dagger$$^\dagger$ & -31.27 & 0.31 & 80.54 & 0.56 \\
 840.597641 (M) & 10.80 & 0.40 & 26.34 & 0.90 \\
 842.601830 (M) & 76.18 & 0.41 & -54.28 & 0.90 \\
 841.625584 (M) & 52.52 & 0.19 & -24.49 & 0.41 \\
 844.616603 (M) & 53.94 & 0.24 & -26.92 & 0.48 \\
 845.691457 (M) & 27.92 & 0.31 & 6.55 & 0.72 \\
 846.687489 (M) & 4.52 & 0.30 & 33.65 & 0.82 \\
 844.665031(L)$^\S$ & 53.16 & 0.16 & -24.99 & 0.49 \\
 846.633214 (L) & 6.15 & 0.23 & 31.05 & 0.59 \\
 843.869932(K)$^\ddagger$ & 69.07 & 0.08 & -46.49 & 0.21 \\
 850.792710 (K) & -21.32 & 0.10 & 68.47 & 0.25 \\
\hline
\multicolumn{5}{l}{$\dagger$: T = Fred L. Whipple Observatory, Tillinghast 1.5m} \\
\multicolumn{5}{l}{$\dagger$$\dagger$: M = McDonald Observatory, Harlan J. Smith Telescope 2.7m} \\
\multicolumn{5}{l}{$\S$: L = Lick Observatory, Shane 3m} \\
\multicolumn{5}{l}{$\ddagger$: K = Keck Observatory} \\
\end{tabular}
\end{center}
\end{table}

\clearpage
\begin{table}[ht]
\begin{center}
\footnotesize
\caption{\kic: the Eclipsing Binary
\label{tab:the_EB}}
\begin{tabular}{l|cccc}
\hline
\hline
Parameter & Value & Uncertainty (1$\sigma$) & Unit & Note \\
\hline
~Orbital Period, $P_{\rm bin}$ & 11.25882 & 0.00060 & days & Pr{\v s}a et al.~(2011) \\
~Epoch of Periastron passage, $T_{\rm 0}$ & -47.86903 & 0.00267 & days$^\dagger$ & Spectroscopy \\
~Velocity semi-amplitude, $K_{\rm A}$ & 55.73 & 0.21 & km s$^{-1}$ & Spectroscopy \\
~Velocity semi-amplitude, $K_{\rm B}$ & 69.13 & 0.50 & km s$^{-1}$ & Spectroscopy \\
~System velocity offset, $\gamma$ & 18.00 & 0.13 & km s$^{-1}$ & Spectroscopy \\
~Eccentricity, $e_{\rm bin}$ & 0.1593 & 0.0030 & & Spectroscopy \\
~Argument of Periapsis, $\omega_{\rm bin}$ & 300.85 & 0.91 & \arcdeg & Spectroscopy\\
~Semi-major Axis, $a_{\rm A}\sin{i}$ & 12.24 & 0.04 & $R_\odot$ & Spectroscopy \\
~Semi-major Axis, $a_{\rm B}\sin{i}$ & 15.19 & 0.11 & $R_\odot$ & Spectroscopy \\
~Semi-major Axis, $a_{\rm bin}\sin{i}$ & 27.43 & 0.12 & $R_\odot$ & Spectroscopy \\
~Mass of star A, $M_{\rm A}\sin^3{i}$ &1.210 & 0.019 & $M_\odot$ & Spectroscopy \\
~Mass of star B, $M_{\rm B}\sin^3{i}$ & 0.975 & 0.010 & $M_\odot$ & Spectroscopy \\
~Mass ratio, $Q = M_{\rm B}/M_{\rm A}$ & 0.8062 & 0.0063 & & Spectroscopy\\
~Temperature of Star A, $T_{\rm A}$ & 6210 & 100 & K & Spectroscopy \\
~Temperature of Star B, $T_{\rm B}$ & 5770 & 125 & K & Spectroscopy \\
~Flux ratio, $F_{\rm B}/F_{\rm A}$ & 0.21 & 0.02 & & Spectroscopy \\
~V sin i of Star A, $V \sin i$ & 8.4 & 0.5 & km s$^{-1}$ & Spectroscopy \\
~V sin i of Star B, $V \sin i$ & 5.1 & 1.0 & km s$^{-1}$ & Spectroscopy \\
~Fe/H of Star A, $[Fe/H]$ & -0.14 & 0.05 & & Spectroscopy \\
~Age & 4.4 & 0.25 & Gyr & Spectroscopy \\
~Normalized Semi-major Axis, $R_{\rm A}/a_{\rm bin}$ & 0.0655 & 0.0002 & & Photometry$^\dagger$$^\dagger$ \\
~Normalized Semi-major Axis, $R_{\rm B}/a_{\rm bin}$ & 0.0354 & 0.0001 & & Photometry \\
~Flux ratio, $F_{\rm B}/F_{\rm A}$ & 0.22 & 0.02 & & Photometry \\
~Flux Contamination $F_{\rm cont, imaging}$ & 0.069 & 0.015 & & Photometry \\
~Flux Contamination (mean), $F_{\rm cont, MAST}$ & 0.04 & 0.01 & & MAST \\
\hline
\hline
\multicolumn{4}{l}{$\dagger$: BJD - 2,455,000} \\
\multicolumn{4}{l}{$\dagger$$\dagger$: Based on pre-photodynamic analysis of the \kepler data.} \\
\end{tabular}
\end{center}
\end{table}

\clearpage
{\renewcommand{\arraystretch}{1.45}%
\begin{table}[ht]
\scriptsize
\begin{center}
\caption{Fitting parameters for the ELC photodynamical solution of the \kic system.
\label{tab:ELC_in}}
\begin{tabular}{c|ccc}
\hline
\hline
~~Parameter & Mode (with $68\%$-range) & Mean (with $68\%$-range) & Unit \\
\hline
 ~{\bf Binary Star} & & & \\
~~Orbital Period, $P_{\rm bin}$ & $11.2588185^{+0.0000009}_{-0.0000007}$ & $11.2588186^{+0.0000008}_{-0.0000007}$ & days \\
~~Time of Conjunction, $T_{\rm conj}$ & $-43.51995^{+0.00002}_{-0.000001}$ & $-43.5199517^{+0.00001}_{-0.000001}$ & BJD-2,455,000 \\
~~$e_{\rm b}\sin\omega_{\rm b}$ & $-0.1386^{+0.0008}_{-0.0005}$ & $-0.1384^{+0.0007}_{-0.0007}$ & \\
~~$e_{\rm b}\cos\omega_{\rm b}$ & $0.081418^{+0.000009}_{-0.000008}$ & $0.081419^{+0.000008}_{-0.000009}$ & \\
~~Inclination, $i_{\rm bin}$ & $87.9305^{+0.0143}_{-0.0185}$ & $87.9271^{+0.0176}_{-0.0151}$ & \arcdeg \\
~~Mass ratio, $Q = M_{\rm B}/M_{\rm A}$ & $0.7943^{+0.0016}_{-0.0023}$ & $0.7939^{+0.0019}_{-0.0019}$ & \\
~~Velocity semi-amplitude, $K_{\rm A}$ & $55.2159^{+0.0544}_{-0.0665}$ & $55.2091^{+0.0611}_{-0.0597}$ & km s$^{-1}$ \\
~~Fractional Radius, $R_{\rm A}/R_{\rm B}$ & $1.8562^{+0.0038}_{-0.0071}$ & $1.8545^{+0.0055}_{-0.0054}$ & \\
~~Temperature ratio, $T_{\rm B}/T_{\rm A}$ & $0.9360^{+0.0026}_{-0.0019}$ & $0.9363^{+0.0024}_{-0.0022}$ & \\
~~Limb darkening  primary \kepler, $x1_{\rm I}$ & $0.9825^{+0.015}_{-0.49}$ & $0.6641^{+0.3333}_{-0.1717}$ & \\
~~Limb darkening  primary \kepler, $x1_{\rm J}$ & $0.3975^{+0.2050}_{-0.14}$ & $0.4968^{+0.1057}_{-0.2393}$ & \\ 
~~Limb darkening secondary \kepler, $x1_{\rm U}$ & $0.2708^{+0.044}_{-0.0231}$ & $0.2847^{+0.03}_{-0.0371}$ & \\
~~Limb darkening secondary \kepler, $y1_{\rm U}$ & $0.361^{+0.06}_{-0.054}$ & $0.3701^{+0.0509}_{-0.0631}$ & \\
~~Apsidal constant, $k_{2}(\rm A)$ & $0.0062^{+0.0023}_{-0.0030}$ & $0.0058^{+0.0026}_{-0.0027}$ & \\                   
~~Apsidal constant, $k_{2}(\rm B)$$^\dagger$: & 0.0-0.03 & -- & \\ 
~~Rotational-to-orbital frequency ratio, primary  & $1.0751^{+0.0324}_{-0.0324}$ & $1.0759^{+0.0315}_{-0.0333}$ & \\
~~Rotational-to-orbital frequency ratio, secondary & $1.3509^{+0.17325}_{-0.2153}$ & $1.3152^{+0.2089}_{-0.1796}$ & \\
\hline
~{\bf CB Planet} & & & \\
~~Orbital Period, $P_{\rm p}$ &  $1107.5946^{+0.0173}_{-0.0119}$ & $1107.6056^{+0.0063}_{-0.0229}$ & days \\
~~Time of Conjunction, $T_{\rm conj}$ & $-1.5005^{+0.0058}_{-0.0058}$ & $-1.5005^{+0.0059}_{-0.0058}$ & BJD-2,455,000 \\
~~$e_{\rm p}\sin\omega_{\rm p}$ & $0.02516^{+0.0025}_{-0.0057}$ & $0.02063^{+0.0069}_{-0.0012}$ & \\
~~$e_{\rm p}\cos\omega_{\rm p}$ & $-0.0006^{+0.1232}_{-0.0784}$ & $0.0394^{+0.0832}_{-0.1184}$ & \\
~~Inclination , $i_{\rm p}$ & $90.0962^{+0.0026}_{-0.0036}$ & $90.0945^{+0.00436}_{-0.0019}$ & \arcdeg \\
~~Nodal Longitude, $\Omega_{\rm p}$ & $-2.0991^{+0.2041}_{-0.3368}$ & $-2.3341^{+0.4392}_{-0.1017}$ & \arcdeg \\
~~Mass of Planet b, $M_{\rm p}$ & $312.5000^{+175.0000}_{-308.3333}$ & $344.4662^{+143.0338}_{-340.2996}$ &  $M_{\rm Earth}$ \\
\hline
~{\bf Seasonal Contamination} & & & \\
~~Season 0 & $0.0776^{+0.0063}_{-0.0057}$ & $0.0779^{+0.0059}_{-0.0061}$ & \\
~~Season 1 & $0.0670^{+0.0066}_{-0.0054}$ & $0.0676^{+0.0060}_{-0.0060}$ &  \\
~~ Season 2 & $0.0798\pm0.0024$ & & \\
~~Season 3 & $0.0753^{+0.0060}_{-0.0060}$ & $0.0751^{+0.0062}_{-0.0058}$ & \\
\hline
\multicolumn{4}{l}{$^\dagger$: Allowed range.} \\
\end{tabular}
\end{center}
\end{table}}

\clearpage
{\renewcommand{\arraystretch}{1.55}%
\begin{table}[ht]
\scriptsize
\begin{center}
\caption{Photodynamically-derived parameters for the \kic system (osculating at BJD=2,455,000). The best-fit column reproduces the light-curve; the mode and mean columns represent the MCMC-optimized parameters.  
\label{tab:ELC_out}}
\begin{tabular}{ccccc}
\hline
\hline
~~Parameter & Best-fit & Mode (with $68\%$-range) & Mean (with $68\%$-range) & Unit \\
\hline
~~Mass of Star A, $M_{\rm A}$ & $\vMstarA\pm\eMstarA$ & $1.2167^{+0.0054}_{-0.0059}$ & $1.2163^{+0.0059}_{-0.0054}$ & $M_\odot$ \\
~~Mass of Star B, $M_{\rm B}$ & $\vMstarB\pm\eMstarB$ & $0.9652^{+0.0036}_{-0.0027}$ & $0.9656^{+0.0031}_{-0.0031}$ & $M_\odot$ \\
~~Mass of Planet b, $M_{\rm p}$$^\dagger$ & $\vMplanetear\pm\eMplanetear$ & $312.5000^{+175.0000}_{-308.3333}$ & $344.4662^{+143.0338}_{-340.2996}$ & $M_{\rm Earth}$ \\
~~Radius of Star A, $R_{\rm A}$ & $\vRstarA\pm\eRstarA$ & $1.7871^{+0.0026}_{-0.0043}$ & $1.7873^{+0.0034}_{-0.00352}$ & $R_\odot$ \\
~~Radius of Star B, $R_{\rm B}$ & $\vRstarB\pm\eRstarB$ & $0.9636^{+0.0032}_{-0.0030}$ & $0.9637^{+0.0030}_{-0.0031}$ & $R_\odot$ \\
~~Radius of Planet, $R_{\rm p}$ & $\vRplanetear\pm\eRplanetear$ & $11.8504^{+0.0677}_{-0.0804}$ & $11.8438^{+0.0743}_{-0.0738}$ & $R_{\rm Earth}$ \\
~~Gravity of Star A, $\log g_{\rm A}$ & $4.0180\pm0.0020$ & $4.0181^{+0.0017}_{-0.0016}$ & $4.0182^{+0.0016}_{-0.0017}$ & cgs \\
~~Gravity of Star B, $\log g_{\rm B}$ & $4.4534\pm0.0040$ & $4.4555^{+0.0018}_{-0.0032}$ & $4.4544^{+0.0028}_{-0.0022}$ & cgs \\
\hline
~{\bf Binary Orbit} & & \\
~~Orbital Period, $P_{\rm bin}$$^\dagger$ & $11.2588179\pm0.0000013$ & $11.2588185^{+0.0000009}_{-0.0000007}$ & $11.2588186^{+0.0000008}_{-0.0000007}$ & days \\
~~Time of Conjunction, $T_{\rm conj}$$^\dagger$ & $-43.51995\pm0.00002$ & $-43.51995^{+0.00002}_{-0.000001}$ & $-43.5199517^{+0.00001}_{-0.000001}$ & BJD-2,455,000 \\
~~Semimajor Axis, $a_{\rm bin}$ & $\vabin\pm\eabin$ & $0.1275^{+0.0002}_{-0.0002}$ & $0.12751^{+0.0002}_{-0.0002}$ & AU \\
~~Eccentricity, $e_{\rm bin}$ & $\vebin\pm\eebin$ & $0.1607^{+0.0005}_{-0.0006}$ & $0.1606^{+0.0006}_{-0.0006}$ & \\
~~Inclination, $i_{\rm bin}$$^\dagger$ & $87.9164\pm0.0145$ & $87.9305^{+0.0143}_{-0.0185}$ & $87.9271^{+0.0176}_{-0.0151}$ & \arcdeg \\
~~Argument of Periastron, $\omega_{\rm bin}$ & $300.5442\pm0.0883$ & $300.4233^{+0.1600}_{-0.0867}$ & $300.4621^{+0.1213}_{-0.1254}$ & \arcdeg \\
~~Apsidal Precession, $\Delta\omega$ (ELC)$^\ddagger$ & 0.0002420 & -- & -- & $^{\circ}/\rm cycle$ \\
~~Apsidal Precession, $\Delta\omega$ (analytic), GR & 0.0001873 & -- & -- & $^{\circ}/\rm cycle$ \\
~~Apsidal Precession, $\Delta\omega$ (analytic), tidal & 0.0000336 & -- & -- & $^{\circ}/\rm cycle$ \\
\hline
~{\bf Planetary Orbit} & & & \\
~~Orbital Period, $P_{\rm p}$$^\dagger$ & $1107.5923\pm0.0227$ & $1107.5946^{+0.0173}_{-0.0119}$ & $1107.6056^{+0.0063}_{-0.0229}$ & days \\
~~Time of Conjunction, $T_{\rm conj}$$^\dagger$ & $-1.5028\pm0.0049$ & $-1.5005^{+0.0058}_{-0.0058}$ & $-1.5005^{+0.0059}_{-0.0058}$ & BJD-2,455,000 \\
~~Semimajor Axis, $a_{\rm p}$ & $\vaplanet\pm\eaplanet$ & $2.7183^{+0.0032}_{-0.0040}$ & $2.7177^{+0.0038}_{-0.0034}$ & AU \\
~~Eccentricity , $e_{\rm p}$ & $0.0581\pm0.0689$ & $0.0275^{+0.08165}_{-0.0035}$ & $0.0881^{+0.0210}_{-0.0641}$ & \\
~~Inclination , $i_{\rm p}$$^\dagger$ & $90.0972\pm0.0035$ & $90.0962^{+0.0026}_{-0.0036}$ & $90.0945^{+0.00436}_{-0.0019}$ & \arcdeg \\
~~Argument of Periastron, $\omega_{\rm p}$ & $155.0464\pm146.5723$ & $4.250^{+161.5667}_{-4.933}$ & $68.7878^{+97.0289}_{-69.4711}$ & \arcdeg \\
~~Nodal Longitude, $\Omega_{\rm p}$$^\dagger$ & $-2.0393\pm0.3643$ & $-2.0991^{+0.2041}_{-0.3368}$ & $-2.3341^{+0.4392}_{-0.1017}$ & \arcdeg \\
~~Mutual Orbital Inclination, $\Delta i$$^\dagger$$^\dagger$ & $2.9855\pm0.2520$ & $3.019^{+0.238}_{-0.140}$ & $3.194^{+0.062}_{-0.316}$ & \arcdeg \\
\hline
\multicolumn{5}{l}{$^\dagger$: For easier interpretation, we repeat here the Mode and Mean for these parameters that are listed in Table~\ref{tab:ELC_in} as well.} \\
\multicolumn{5}{l}{\bf $^\ddagger$: $k_{2}(\rm A) = 0.00249\pm0.00522$ and $k_{2}(\rm B) = 0.02979\pm0.00053$.} \\
\multicolumn{5}{l}{$^\dagger$$^\dagger$: $\cos(\Delta i) = \sin(i_{\rm bin})\sin(i_{\rm p})\cos(\Delta\Omega) + \cos(i_{\rm bin})\cos(i_{\rm p})$} \\
\end{tabular}
\end{center}
\end{table}}

\begin{table}[ht]
\begin{center}
\caption{Best-fit, barycentric Cartesian coordinates for the \kic system at BJD=2,455,000. The observer is looking along the $+z$-direction, and the sky is in the $xy$ plane.
\label{tab:Barycentric_cartesian}}
\scriptsize
\begin{tabular}{c|ccc}
\hline
\hline
Parameter & Primary Star & Secondary Star & CB Planet \\
\hline
$M$ [$M_\odot$] & 1.22067659415220042 & 0.967766001496474848 & 1.45159791061409129$\times10^{-3}$ \\
$x$ [AU] & -3.78722501644566112$\times10^{-2}$ & 4.78054676980426904$\times10^{-2}$ & -2.39301668324773467$\times10^{-2}$ \\
$y$ [AU] & -1.59238464640225146$\times10^{-3}$ & 2.01398578685853605$\times10^{-3}$ & -3.63758119907132233$\times10^{-3}$ \\
$z$ [AU] & -4.55884242496607597$\times10^{-2}$ & 5.35371764610109713$\times10^{-2}$ & 2.64347531964575655 \\
$V_x$ [AU/day] & 1.97567794427114841$\times10^{-2}$ & -2.48961816644120461$\times10^{-2}$ & -1.58170933420904401$\times10^{-2}$ \\
$V_y$ [AU/day] & -8.53253808750116679$\times10^{-4}$ & 1.07539120444087718$\times10^{-3}$ & 5.64830825578074218$\times10^{-4}$ \\
$V_z$ [AU/day] & -2.34418182175292165$\times10^{-2}$ & 2.95694015193574272$\times10^{-2}$ & -9.52503798501482327$\times10^{-4}$ \\
\hline
\hline
\end{tabular}
\end{center}
\end{table}

\begin{table}[ht]
\begin{center}
\scriptsize
\caption{Input parameters (osculating) for \kic needed for the {\it photodynam} code (Carter et al.~2011). For details see description at https://github.com/dfm/photodynam .
\label{tab:pd_in}}
\begin{tabular}{llllll}
\hline
\hline
3 & 0.0 & & & & \\
0.02   & 1e-20 & & & & \\
0.00036121310659 & 0.00028637377462 & 0.00000042954554 & & & \\
0.00832910409711 & 0.00449544901109 & 0.00050567865570 & & & \\
0.75809801560924 & 0.16689760802955 & 0.0 & & & \\
0.38919885445149 & 0.46863637334167 & 0.0 & & & \\
0.14297457592624 & 0.02356800780666 & 0.0 & & & \\
0.12763642607906 & 0.16022073112714 & 1.53443053940587 & -1.03769914810282 & 0.0 & 1.581592306154 \\
2.72058352472999 & 0.05807789749374 & 1.57249365140452 & 2.70607009751131 & -0.03559283766606 & -1.023387804660 \\
\hline
\end{tabular}
\end{center}
\end{table}

\begin{table}[ht]
\begin{center}
\scriptsize
\caption{Mid-transit times, depths and durations of the planetary transits.
\label{tab:CBP_transits}}
\begin{tabular}{ccccccc||ccc}
\hline
\hline
 Event \# & Center & $\sigma$ & Depth$^\dagger$ & $\sigma$ & Duration & $\sigma$ & Center & Depth & Duration \\
 & (Time-2455000 [BJD]) & (Center) & [ppm] & (Depth) & [days] & (Duration) & (Time-2455000 [BJD]) & [ppm] & [days] \\
\hline
\hline
{\bf Observed} & & & & & & & {\bf Predicted} & \\
\hline
1 & -3.0018 & 0.0027 & 2070 & 150 & 0.1352 & 0.0125 & -3.0035 & 2080 & 0.1278 \\
2 & 1104.9510 & 0.0041 & 2990 & 250 & 0.4013 & 0.0071 & 1104.9515 & 3210 & 0.3973 \\
3 & 1109.2612 & 0.0036 & 2470 & 70 & 0.2790 & 0.0088 & 1109.2645 & 2450 & 0.2727 \\
\hline
{\bf Future} & & & & & & \\
\hline
4$^\dagger$$^\dagger$ & -- & -- & -- & -- & -- & -- & 2209.5440 & 3200 & 0.4250 \\
5$^\dagger$$^\dagger$ & -- & -- &  & -- & -- & -- & 2213.7253 & 2460 & 0.1756 \\
6$^\ddagger$ & -- & -- & -- & -- & -- & -- & 3314.5362 & 3000 & 0.6597 \\
7$^\ddagger$ & -- & -- & -- & -- & -- & -- & 3317.4837 & 2860 & 0.6840 \\
8$^\ddagger$ & -- & -- & -- & -- & -- & -- & 3317.9441 & 2390 & 0.1517 \\
9$^\ddagger$ & -- & -- & -- & -- & -- & -- & 3321.0339 & 3290 & 0.5604 \\
\hline
\multicolumn{4}{l}{$\dagger$: in terms of $(\frac{r_{\rm p}}{r_{\rm A}})^2$} \\
\multicolumn{4}{l}{$\dagger$$\dagger$: Unsuccessful attempts to observe from the ground.} \\
\multicolumn{4}{l}{$\ddagger$: Within the operation timeframe of the {\it TESS} mission.}
\end{tabular}
\end{center}
\end{table}

\end{document}